\documentclass[prb,aps,nobibnotes,twocolumn]{revtex4}
\usepackage{graphicx}%
\usepackage{dcolumn}
\usepackage{amsmath}   
\usepackage{bm}
\usepackage{dcolumn}
\usepackage{longtable}
\begin{document}

\title{\centering\Large\bf 
Two-Gaussian excitations model for the glass transition} 
\author{Dmitry V.\ Matyushov}
\email[E-mail:]{dmitrym@asu.edu.}
\author{C.\ A.\ Angell} 
\email[E-mail:]{caangell@asu.edu.}  
\affiliation{
  Department of Chemistry and Biochemistry, Arizona State University, PO Box
  871604, Tempe, AZ 85287-1604}
\date{\today}
\begin{abstract}
  We develop a modified ``two-state'' model with Gaussian widths for
  the site energies of both ground and excited states, consistent with
  expectations for a disordered system. The thermodynamic properties
  of the system are analyzed in configuration space and found to
  bridge the gap between simple two state models (``logarithmic''
  model in configuration space) and the random energy model
  (``Gaussian'' model in configuration space).  The Kauzmann
  singularity given by the random energy model remains for very
  fragile liquids but is suppressed or eliminated for stronger
  liquids. The sharp form of constant volume heat capacity found by
  recent simulations for binary mixed Lennard Jones and soft sphere
  systems is reproduced by the model, as is the excess entropy and
  heat capacity of a variety of laboratory systems, strong and
  fragile. The ideal glass in all cases has a narrow Gaussian, almost
  invariant among molecular and atomic glassformers, while the excited
  state Gaussian depends on the system and its width plays a role in
  the thermodynamic fragility. The model predicts the existence of
  first-order phase transition for fragile liquids. The analysis of
  laboratory data for toluene and $o$-terphenyl indicates that fragile
  liquids resolve the Kauzmann paradox by a first-order transition
  from supercooled liquid to ideal glass state at a temperature
  between $T_g$ and Kauzmann temperature extrapolated from
  experimental data.  We stress the importance of the temperature
  dependence of the energy landscape, predicted by the
  fluctuation-dissipation theorem, in analyzing the liquid
  thermodynamics.
\end{abstract}
\preprint{Submitted to J.\ Chem.\ Phys.\ }
\maketitle

\section{Introduction}
\label{sec:1}
In the search for understanding of the glass transition phenomenon,
attention has been focused overwhelming on the dynamic aspects of the
behavior of supercooling
liquids.\cite{Williams:55,Cohen:59,Turnbull:61,Adam:65,Cohen:79,Cohen:81,Frederickson:84,Goetze:91,Ngai:93,Pitts:00,Perera:99,Jung:04,Cang:03,Glotzer:99}
This is natural in view of the general agreement that it is the
falling out of equilibrium, at a temperature that depends on the
cooling rate, which provokes the observed ``drop'' in heat capacity at
$T_g$.  In other words the glass transition phenomenon observed experimentally is an
entirely kinetic phenomenon. However, this approach leaves unresolved
a basic thermodynamic question that has troubled glass scientists for
the best part of a century.

The thermodynamic problem concerns the course of the entropy in excess
of that of the crystal (or any other state whose entropy vanishes at 0
K) during cooling of the equilibrated liquid state. First posed in
1930 by Simon for the particular case of glycerol,\cite{Simon:30} and
after for a variety of substances by Kauzmann,\cite{Kauzmann:48} the
question concerns what physical process occurs to avoid the liquid
entropy intersecting that of the crystal, as simple extrapolation of
the observed entropy changes with decreasing temperature would require
for all fragile liquids.\cite{Kauzmann:48} Unless it can be shown
generally that the liquid becomes mechanically unstable during
cooling, (hence has no option but to crystallize), the resolution of
this problem requires a thermodynamic description of the liquid
entropy which is independent of equilibration time scales. A
mechanical instability due to the vanishing of the nucleation barrier
was Kauzmann's resolution\cite{Kauzmann:48} of what has become known
as the Kauzmann paradox (kinetic phenomenon, $T_g$, avoiding a
thermodynamic crisis, at $T_K$). Although this resolution has been
given recent support from certain crystallizable spin-glass model
studies,\cite{Cavagna:03} there is a broad belief that the Kauzmann
paradox demands a more general resolution.

While there have been a number of insightful investigations of the
thermodynamic properties of glassformers, using the configuration
space energy landscape approach,\cite{Debenedetti:99,Nave:02,Shell:03}
there have been surprisingly few attempts to provide theoretical
functions to describe the liquid thermodynamics in terms of underlying
models.  Early attempts were focused on polymers for which
quasilattice models were plausible.  Considering the case of atactic
polymers, for which no low energy crystalline state exists, Gibbs and
Dimarzio\cite{Gibbs:58} argued that a thermodynamic (equilibrium)
transition of second order, at which the configurational entropy
vanishes, must set the limit to supercooling of the liquid state of
the polymer. It has been broadly supposed that a similar transition
might apply to liquids\cite{Angell:97} though, without the polymer
basis, there is so far less theoretical justification for this.
Furthermore, Stillinger\cite{Stillinger:88} has argued that such a
transition is not possible in principle, though how closely such a
transition could be approached has not been discussed. By contrast, a
free volume model by Cohen and Grest\cite{Cohen:79} has suggested
that, ideally, the transition to the ground state glass should be of
first order, though no experimental example has been identified. On
the other hand, spin models,\cite{Frederickson:84} and their
application to coarse-grained models of dynamics in structural
glasses,\cite{Garrahan:02,Garrahan:03} have treated the thermodynamic
component of the problem as trivial, to be resolved by the
thermodynamics of uncorrelated excitations.

Indeed, it has been long known that simple uncorrelated excitation (or
defect) models of amorphous
solids,\cite{Macedo:66,Angell:72,Perez:85,Angell:00} can give a good
account of the entropy-temperature
relation\cite{Angell:00,Moynihan:00} particularly in elemental cases
like selenium.\cite{Angell:00} These show that the Kauzmann limit
paradox can result as a consequence of an unjustified extrapolation of
the entropy vs temperature relation, which should be continuous,
though rapidly varying, in the vicinity of the Kauzmann temperature.
Unfortunately, excitation models in their usual forms (in which the
excitations are presumed to be non-interacting), predict the
occurrence of a heat capacity maximum above
$T_g$,\cite{Angell:72,Angell:00} which in practice is only found in
some strong liquids.\cite{Angell:77,Hemmati:01}

The simple two state model has recently reappeared under a new name,
the ``logarithmic'' model based on its properties in configuration
space.\cite{Debenedetti:03} Like its real space predecessors, the
logarithmic model has the problem of predicting a heat capacity
maximum where none is found (though when combined in configuration
space with a Gaussian component, this problem is
avoided,\cite{Debenedetti:03} see below). In a variant of such models,
Tanaka\cite{Tanaka:98,Tanaka1:99} has introduced a two order
parameter Landau model for the thermodynamics of glassformers where
bond length and orientation are distinguished.

A defect model with behavior much like that to be described in this
paper (despite a quite different starting point) is the interstitialcy
model of Granato.\cite{Granato:92,Granato:02} This model posits a single
entropy-rich defect (the interstitial defect of crystalline metals)
and obtains the temperature dependence of the defect concentration
from the temperature dependence of the shear modulus. The
cooperativity missing from earlier two state models, or included
\textit{ad hoc},\cite{Angell:72,Angell:00} is built in through the
proportionality of the defect energy to the shear modulus.  The latter
decreases strongly with temperature, leading to laboratory-like heat
capacities and a phase transition at lower temperatures -- which is
assigned to a return to the crystal state.  Alternatively, Wolynes and
co-workers\cite{Xia:00,Lubchenko:04} have described a mosaic model in
which the inter-domain boundary energies play a vital role in the
thermodynamics. With the appropriate assumptions, this model can
resolve the Kauzmann paradox in the same way as does the random energy
model,\cite{Derrida:80,Derrida:81,Richert:90} the system simply
running out of states at a singular (Kauzmann) temperature. The
sudden, latent heat-free, transition to the ground state is described
as a ``random first order'' transition,\cite{Xia:00,Lubchenko:04} the
latent heat of the normal first order transition having been given up
continuously over the supercooling temperature range. Both
mosaic\cite{Xia:00,Lubchenko:04} and constrained
excitation\cite{Garrahan:02,Garrahan:03,Jung:04} models prove capable
of predicting important dynamic features of glassformers, such as the
decoupling of viscosity from diffusivity on approach to the glass
transition from above.\cite{Swallen:03} However, in this paper we are
concerned only with the thermodynamic problem.

Most theoretical models now gain their support from molecular dynamics
computer simulations but, because of their time scale limitations,
these cannot be expected to help much with the long time aspects of
glass transition problem. The Gaussian distributions of
configurational states found in several
cases\cite{Speedy:88,Buchner:99,Sciortino:00,Heuer:00,Sastry:01,Yan:04}
in the shorter relaxation time domain (which, however, covers most of
the inherent structure energy range between the extrapolated Kauzmann
temperature and the high temperature limit) would imply the existence,
at lower temperatures, of a Kauzmann-like singularity. Clearly
something has to change between the lowest temperature of these
simulations and the vanishing entropy temperature, if Stillinger's
argument is to be upheld. In the laboratory behavior of the closest
relatives of the most simulated system, binary mixed Lennard-Jones
(LJ),\cite{Buchner:99,Sciortino:00,Sastry:01} what changes is the
state of the system: it crystallizes, leaving the problem unresolved.

Recent simulation of the thermodynamic behavior of a small periodic
box of the mixed soft sphere system by Grigera and
Parisi,\cite{Grigera:01} Yu and Carruzzo,\cite{Yu:02,Yu:04} and De
Pablo and co-workers,\cite{Yan:04} suggest, however, that if
crystallization does not occur, and equilibrium is maintained, then the
heat capacity continues to increase.  Finally, it peaks sharply and
decreases to zero, like a narrowly avoided Kauzmann singularity.  Yan
\textit{et al.}\cite{Yan:04} suppose that all possible states of the
system have been explored, though this is not yet proven (simulations
by Yu and Carruzzo\cite{Yu:02,Yu:04} actually indicate that heat
capacity drops due to insufficient sampling). In a separate study by
Debenedetti and Stillinger,\cite{Debenedetti:03} a range of behavior
intermediate between the simple two state model and the singularity of
the random energy model\cite{Derrida:81} has been illustrated by
adopting, \textit{ad hoc}, an additive mixture of two-state
(logarithmic model) and Gaussian (random energy model) distributions.
The behavior seen by Yan \textit{et al.},\cite{Yan:04} and required by
experiment,\cite{Angell:72} is found for Gaussian-rich mixtures.

\begin{widetext}
\begin{figure*}
  \centering \includegraphics*[width=15cm]{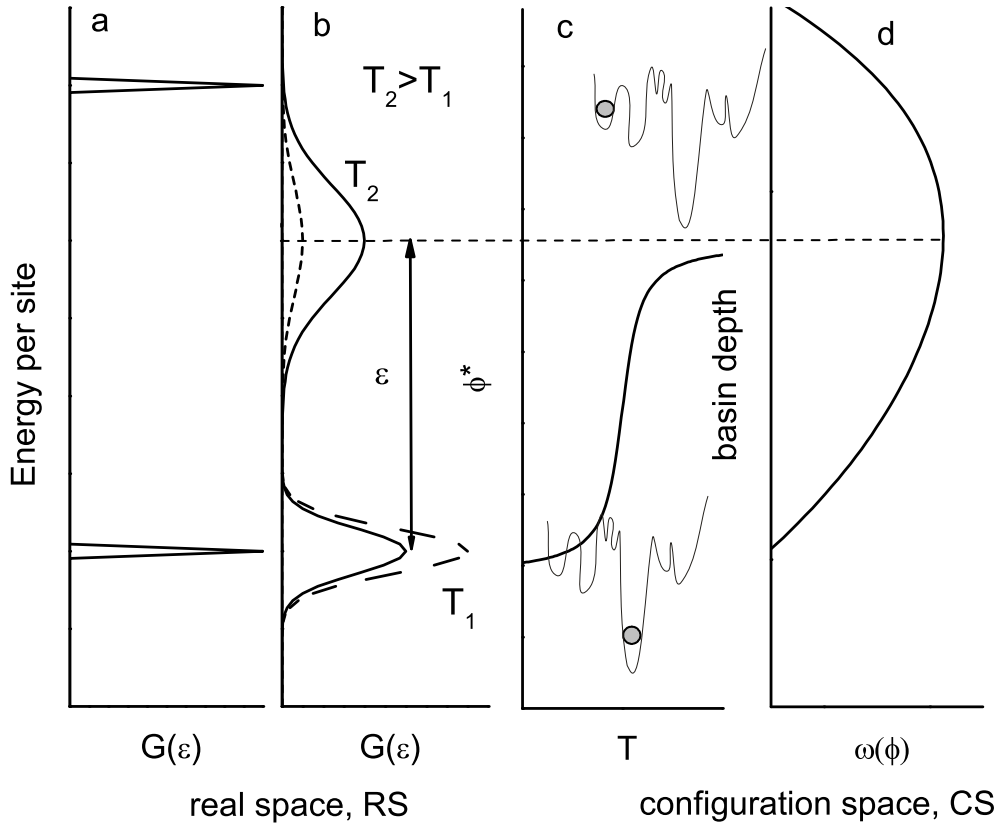}
  \caption{Real space (RS) energies of perfect crystal sites and defect sites (a). 
    Real space energies and populations of the ideal glass and liquid
    sites at temperatures $T_1=T_g$ and $T_2$ respectively (b).
    Configuration space (CS) location of system points at $T_1=T_K$
    and $T_2$, and the excitation profile (sigmoid) (c).
    Quasi-Gaussian enumeration function for configurational states
    (d). }
  \label{fig:1}
\end{figure*}
\end{widetext}

In the present paper we show how thermodynamic behavior of the sort
obtained on soft sphere mixtures\cite{Grigera:01,Yu:02,Yu:04,Yan:04}
can be reproduced by a modified version of the simple excitations
model in which the single (or few) excitation energy(ies) of the
original models is(are) replaced by a more physically reasonable
Gaussian distribution, the centroid of which may lie near but
generally below the value of the original excitation energy, and the
width of which may vary. The existence of such character has been
suggested by analysis of spectral band-shapes for glasses and
liquids\cite{Angell:80} and recently, also, the vibrational density of
states of glasses of different fictive temperatures in which a
quasi-two state behavior is found for the temperature dependence of
the vibrational density of states.\cite{Mossa:02,Angell:03} We note
that the Gaussian analysis that is often used to describe the widths
of spectral bands, is only appropriate if the modes are localized -
which is a poor approximation when dealing with the vibrational
density of states, even for the boson peak.\cite{Gurevich:03} Other
spectroscopic evidence for distinct broken bond excitations in glasses
has been given for weak network liquids\cite{Angell:70} and
recently\cite{Angell:04,com:1} revived in connection with the
boson peak controversy.

\section{Model representations in real space and configuration space (energy landscape).}
\label{sec:2}
The model assumes the presence in the condensed phase of degrees of
freedom which, in real space (RS), can exist in ground (low-energy) and
excited (high-energy) states.\cite{Angell:72} To visualize the model
and its relation to crystal defect physics on the one hand, and to
energy landscape representations on the other, we use Fig.\ \ref{fig:1}. The
distribution of energies in the real space ground state, in the case
of the crystal, is a delta function, on the unit cell length scale,
and a small number of delta functions on the per molecule scale (Fig.\ 
\ref{fig:1}a). The defect states are likewise few in number and well
defined in energy.  In the glass, however, the sites are not all
equivalent on these length scales, and a Gaussian distribution in
energy is expected, both for molecules in the ideal glass and for the
elementary configurational excitation (or defect) states which we
suppose to exist (Fig.\ \ref{fig:1}b).  Excitations may be related to
coordination defects in covalent materials or to local distortions or
packing strains in molecular crystals.

The distribution of energies $\epsilon_i$ in real space is characterized by
two Gaussians $G_i(\epsilon_i)$ where $i=1$ stands for the ground state and
$i=2$ stands for the excited state.  The Gaussian function $G_i(\epsilon_i)$
is defined by the average $\epsilon_{0i}$ and the variance $\sigma_i$. In addition
to the change in energy, the creation of a local defect may result in
an entropy increase related either to a change in the vibrational or
configurational density of states.\cite{Angell:00} The entropy change
per molecule of the glass is $s_0=\Delta S_0/Nk_{\text{B}}$.

The potential energy in configuration space (CS) is a hypersurface
depending on all degrees of freedom of the disordered liquid. The
overall configuration space is decomposed into basins of local
potential energy minima termed inherent
structures.\cite{Stillinger:82,Stillinger:88} The ideal glass, in
configuration space, is represented by the lowest energy basin on the
energy landscape, and any excitation of defect states will lift the
energy to one or other of the higher energy basins. The more defects
in the real space quasilattice, the higher the energy of the
configuration space basin occupied by the system (Fig.\ \ref{fig:1}c).
Thus as the intensity or occupation number of the second real space
Gaussian increases (cf.\ dashed to solid lines in Fig.\ \ref{fig:1}b),
the system point in configuration space moves higher on the landscape
(Fig.\ \ref{fig:1}c).

The distribution of basin energies (CS) is found, by simulation
studies, to conform to a Gaussian (Fig.\ \ref{fig:1}d), but it is
barely possible to distinguish between a Gaussian and the binomial
distribution that is expected for a two-state system, except at the
wings, which are unexplored in any simulations on accessible time
scales. The difference must diminish further for the case where the
excitation energy is distributed, as in our model. Where a given
excitation can occur at any energy in the real space distribution, the
states in the configuration space distribution are only occupied on
the low energy side of the maximum of the Gaussian (Fig.\ 
\ref{fig:1}d). The energies in this half Gaussian, though, are
uniformly higher than those in the crystal manifold, which is very
narrow (Fig.\ \ref{fig:1}c), because the crystal generates very few
defects before it becomes thermodynamically unstable and melts.

\begin{figure}[htbp]
  \centering
  \includegraphics*[width=8cm]{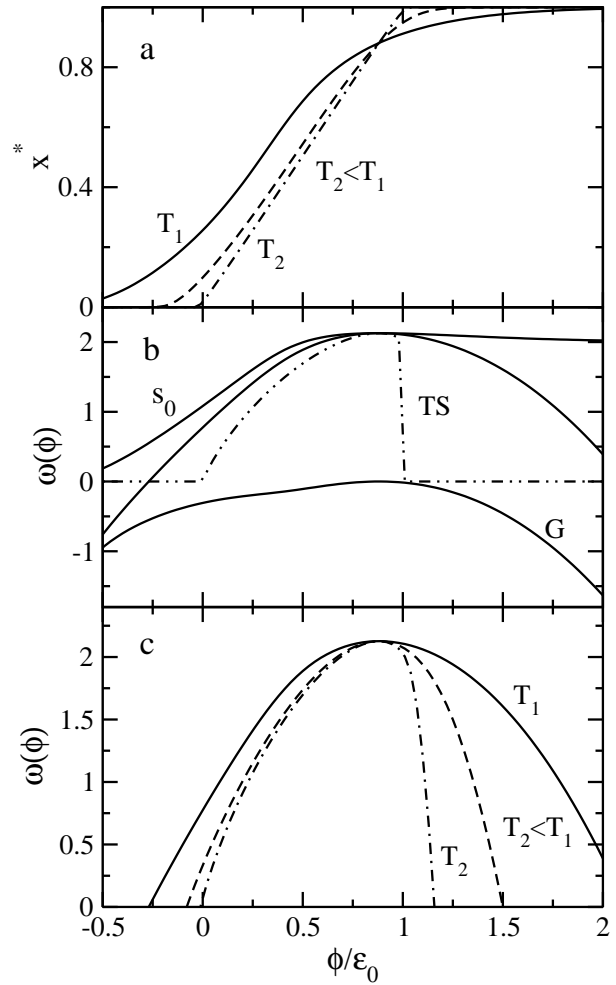}
  \caption{(a): $x^*=x(\phi^*)$ calculates as the root of Eq.\ (\ref{eq:2-18}). 
    (b) $\omega(\phi)$ from Eq.\ (\ref{eq:2-10}) and its separation into the
    ideal mixture term ``$s_0$'' (first summand in Eq.\ 
    (\ref{eq:2-10})) and the Gaussian term ``G'' (second summand in
    Eq.\ (\ref{eq:2-10})). (c): $\omega(\phi)$ from Eq.\ (\ref{eq:2-10}).  The
    lines in (a)--(c) refer to different temperatures: $T= 500$ K
    (solid line), $T=200$ K (dashed line), and $T=10$ K (dash-dotted
    line); $\lambda_2/k_{\text{B}} =200$ K, $\lambda_1/k_{\text{B}} =100$ K,
    $\epsilon_0/k_{\text{B}}=800$ K, $s_0=2.0$. The dash-dotted line in (b)
    shows the enumeration function of the two-state, ``TS'', model
    [Eq.\ (\ref{eq:2-12})]. }
  \label{fig:2}
\end{figure}

The density of inherent structures identified with basins of depth
$\phi$ defines the enumeration function $\omega(\phi)$. Following
Derrida\cite{Derrida:80} and Wolynes,\cite{Bryngelson:87} $\omega(\phi)$ can
be found by summation over all populations of the excited state
\begin{equation}
  \label{eq:2-1}
  e^{N\omega(\phi)}= \sum_{N_2=0}^N C(x) \left[P(\phi,x) \right]^N ,
\end{equation}
where $N_2$ is the number of excitations out of $N$ molecules in the
system, $x= N_2/N$ is the population of the excited state.  The
function $C(x)$ in Eq.\ (\ref{eq:2-1}) is the number of realizations of
a given distribution of molecules between the ground and excited
states
\begin{equation}
  \label{eq:2-2}
  C(x) = \frac{N!}{(N-N_2)!N_2!} e^{s_0N_2}.
\end{equation}
The distribution of basin energies $P(\phi,x)$ in configuration space
[Eq.\ (\ref{eq:2-1})] can be obtained from the real space Gaussians
(Fig.\ \ref{fig:1}b),
\begin{equation}
  \label{eq:2-3}
  P(\phi,x) = \int \delta\left(\phi - x\epsilon_2 - (1-x)\epsilon_1\right) 
              G_2(\epsilon_2) G_1(\epsilon_1) d\epsilon_2 d\epsilon_1. 
\end{equation}
Equation (\ref{eq:2-3}) gives a Gaussian distribution of basin
energies with the average and variance dependent on the population of
excited states in real space
\begin{equation}
  \label{eq:2-4}
   P(\phi,x) = \left[2\pi\sigma\right]^{-1/2} \exp\left[-\frac{(\phi - x\epsilon_0)^2}{2\sigma(x)^2} \right],
\end{equation}
where $\epsilon_0=\epsilon_{02}-\epsilon_{01}$ is the excitation energy and
\begin{equation}
  \label{eq:2-5}
  \sigma(x)^2 = (1-x)^2 \sigma_1^2 + x^2 \sigma_2^2 
\end{equation}
is the variance of basin energies.

In the thermodynamic limit $N\to \infty$, the behavior of the sum in Eq.\ 
(\ref{eq:2-1}) is defined by its largest summand. One finds
\begin{equation}
  \label{eq:2-6}
  \omega(\phi) = s(\phi,x(\phi)),
\end{equation}
where the population at maximum $x(\phi)$ is obtained from the stationary point
\begin{equation}
  \label{eq:2-7}
  \frac{ds(\phi,x)}{dx}\bigg|_{x=x(\phi)} = 0.
\end{equation}
The function $s(\phi,x)=S(\phi,x)/Nk_{\text{B}}$ is the entropy per molecule
at a given population of excited states. From Eqs.\ (\ref{eq:2-1}), (\ref{eq:2-2}),
and (\ref{eq:2-4}) one finds
\begin{equation}
  \label{eq:2-8}
  s(\phi,x) = s_0(x) - \frac{(\phi - x \epsilon_0)^2}{2\sigma(x)^2},
\end{equation}
where $s_0(x)$ is the entropy of an ideal binary mixture
\begin{equation}
  \label{eq:2-9}
  s_0(x) = xs_0 - x\ln x - (1-x)\ln(1-x) .
\end{equation}

The requirement to maximize the entropy at a given basin depth makes
the population $x$ a function of $\phi$ and, therefore, transforms the
enumeration function
\begin{equation}
  \label{eq:2-10}
  \omega(\phi) = s_0(x(\phi)) - \frac{(\phi - x(\phi)\epsilon_0)^2}{2\sigma(x(\phi))^2} 
\end{equation}
into a generally non-Gaussian dependence on $\phi$ (Fig.\ \ref{fig:2}). The
solution $x(\phi)$ is a root of Eqs.\ (\ref{eq:2-7}) and (\ref{eq:2-8}) 
(Fig.\ \ref{fig:2}a). 

In the static energy landscape picture a system at constant volume has
a unique energy landscape in the configuration space that is fixed by
the intermolecular potential for the particles in the system.  From
this viewpoint, it seems natural to assume that the variance $\sigma$ is
independent of temperature. If, in addition, the dependence of $x(\phi)$
and $\sigma(\phi)$ on $\phi$ is neglected in Eq.\ (\ref{eq:2-10}), one arrives at
the standard Gaussian model equivalent to the random energy model
introduced by Derrida.\cite{Derrida:80,Derrida:81} The Derrida model
predicts an ideal glass transition at the Kauzmann temperature
$k_{\text{B}}T_K = \sigma/ \sqrt{2s_0(x(\phi^*))}$ ($\omega(\phi^*)=0$ at $T=T_K$) and a
hyperbolic temperature dependence of the average basin energy
\begin{equation}
  \label{eq:2-11}
  \phi^* = a - b/T.
\end{equation}
The dependence of the type given by Eq.\ (\ref{eq:2-11}) has indeed
been observed in several simulation studies of binary Lennard-Jones
(LJ) mixtures,\cite{Sastry:01,Mossa:02} although deviations from this
law have also been
reported.\cite{Buchner:99,Heuer:00,Voivod:01,Voivod:04,Saksaengwijit:04}
The heat capacity per molecule $c_V =C_V/Nk_{\text{B}}=
k_{\text{B}}^{-1}d\phi^*/dT$ then varies as $c_V \propto 1/T^2$.

\begin{figure}[htbp]
  \centering
  \includegraphics*[width=7cm]{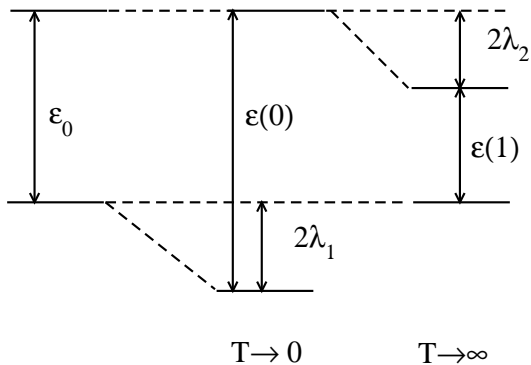}
  \caption{Temperature dependence of the effective excitation energy $\epsilon(x^*)$ [Eq.\ (\ref{eq:2-19})].
    $\epsilon(0)$ and $\epsilon(1)$ denote the excitation energy at zero and
    infinite temperature, respectively. }
  \label{fig:11}
\end{figure}

In the limit when the RS Gaussians are much narrower than the
excitation energy, $\sigma_i \ll \epsilon_0$, one gets the two-state
model.\cite{Angell:72} The Gaussian term in Eq.\ (\ref{eq:2-10}) then
generates a delta function in the density of states
$e^{\omega(\phi)}$ requiring $x(\phi)=\phi/ \epsilon_0$.  This
solution also limits the range of accessible basin depths by the
condition 
\begin{equation}
\label{eq:D}
   0\leq \phi \leq \epsilon_0 . 
\end{equation}
The enumeration function in this limit corresponds to the logarithmic
energy landscape of Debenedetti, Stillinger, and
Shell\cite{Debenedetti:03}
\begin{equation}
  \label{eq:2-12}
  \omega^{\text{TS}}(\phi) = s_0 u  -  u\ln(u) - (1-u) \ln(1-u) ,
\end{equation}
where $u=\phi/ \epsilon_0$, $0\leq u\leq 1$ and the superscript ``TS'' refers to the
two-state model. Note that $s_0=0$ is used in the logarithmic model
in Ref.\ \onlinecite{Debenedetti:03}.

The average basin depth is proportional to the population $x_{\text{TS}}^*$ of
the RS excited states in the two-state limit:
\begin{equation}
  \label{eq:2-13}
  \phi^* = x_{\text{TS}}^*\epsilon_0,\quad x_{\text{TS}}^*=[1+e^{-s_0+\beta\epsilon_0}]^{-1} ,
\end{equation}
$\beta=1/k_{\text{B}}T$. The constant volume heat capacity per particle
(in $k_{\text{B}}$ units) is of Schottky's
form\cite{Moynihan:00,Odagaki:02}
\begin{equation}
  \label{eq:2-14}
  c_V^{\text{TS}}  = (\beta\epsilon_0)^2x_{\text{TS}}^*(1-x_{\text{TS}}^*) .
\end{equation}
This heat capacity form is continuous down to zero K hence, as is well
known,\cite{Angell:72,Perez:85,Angell:00,Moynihan:00} the two-state
model eliminates the ideal glass transition ($T_K\to 0$). Thus variation
of the Gaussian width parameters of our model produces the same
systematic change of heat capacity form that Debenedetti \textit{et
al}.\cite{Debenedetti:03} demonstrated by \textit{ad hoc} linear
mixing of the Gaussian and binomial distribution CS functions. The
heat capacity function for laboratory glasses of different fragility
should then reflect the width of the RS Gaussians relative to the
separation of their centers.

The present model, which we will refer to as the two-Gaussian (2G)
model, projects the two RS Gaussians onto CS enumeration function
given by Eq.\ (\ref{eq:2-10}).  $\omega(\phi)$ in Eq.\ (\ref{eq:2-10}) is
formally a linear combination of the ideal mixture entropy of the RS
two-state model and the CS Gaussian term of the Gaussian model. The
two terms are connected through the $x(\phi)$ function.  $x(\phi)$ is not a
linear function of the two-state model [Eq.\ (\ref{eq:2-13})] when
$\sigma_i \neq 0$, although it approaches the linear limit with lowering
temperature (Fig.\ \ref{fig:2}a) when RS Gaussians become narrower
(see Eq.\ (\ref{eq:2-20})).

Both the ideal mixture term $s_0(x)$ (first summand in Eq.\ 
(\ref{eq:2-10})) and the Gaussian term (second summand in Eq.\ 
(\ref{eq:2-10})) are non-parabolic functions of $\phi$. A bell-shaped
enumeration function, which at high temperatures can be approximated
by a Gaussian shape, is a result of combining two non-Gaussian
summands in Eq.\ (\ref{eq:2-10}) (Fig.\ \ref{fig:2}b).  The two-state
limit of the 2G model results in a very asymmetric enumeration
function when the excitation entropy $s_0$ is non-zero due to the
cutoff of the range of accessible basin energies (Eq.\ (\ref{eq:D});
dash-dotted line in Fig.\ \ref{fig:2}b).
 
The connection between the microcanonical entropy $\omega(\phi)$ and the
canonical ensemble, which permits calculations at given temperatures,
can be obtained by use of two thermodynamic relations\cite{Landau5}
\begin{equation}
  \label{eq:2-15}
  \left(\frac{\partial \omega(\phi)}{\partial \phi}\right)_{N,V} = \beta
\end{equation}
and
\begin{equation}
  \label{eq:2-16}
   \left(\frac{\partial^2 \omega(\phi)}{\partial \phi^2}\right)_{N,V} = - \frac{\beta^2}{c_V}.
\end{equation}

\begin{figure}[htbp]
  \centering
  \includegraphics*[width=8cm]{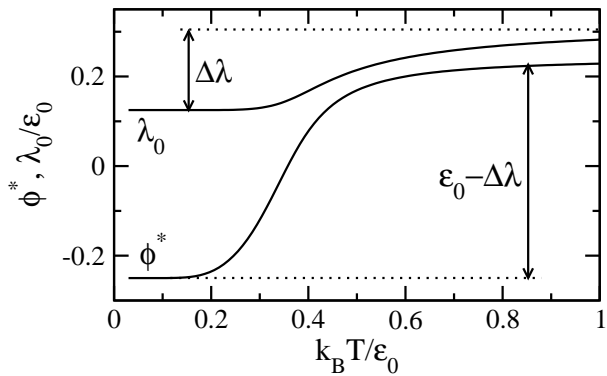}
  \caption{$\lambda_0(T)$ [Eq.\ (\ref{eq:2-21-1})] and $\phi^*$ 
           [Eqs.\ (\ref{eq:2-17}) and (\ref{eq:2-18})] vs temperature; 
           $\lambda_1/k_{\text{B}}=100$ K, $\lambda_2/k_{\text{B}}=200$ K, 
           $\epsilon_0/k_{\text{B}}=800$ K, $s_0=2.0$. }
  \label{fig:3}
\end{figure}

Equation (\ref{eq:2-15}) leads to the average energy of inherent
structures
\begin{equation}
  \label{eq:2-17}
  \phi^* = x^* \epsilon_0 - 2\lambda_0 ,
\end{equation}
where $x^* = x(\phi^*)$ and
\begin{equation}
\label{eq:2-21-1}
    \lambda_0 =(1-x^*)^2 \lambda_1 + (x^*)^2\lambda_2 .
\end{equation}
The coupling parameters $\lambda_i$ in Eq.\ (\ref{eq:2-21-1}) are defined in
terms of the RS distribution widths as
\begin{equation}
  \label{eq:2-20}
  \sigma_i^2 = 2k_{\text{B}}T\lambda_i .
\end{equation}
Equations (\ref{eq:2-5}) and (\ref{eq:2-21-1}) immediately lead to Eq.\ 
(\ref{eq:2-13}) in the limit of narrow RS Gaussians, $\lambda_i\to 0$.
Once Eq.\ (\ref{eq:2-17}) is substituted into Eq.\ (\ref{eq:2-7}), one
arrives at a single, self-consistent equation for $x^*$ which is used to obtain the
average energy $\phi^*$ in Eq.\ (\ref{eq:2-17}):
\begin{equation}
  \label{eq:2-18}
  x^* = \left[1 + e^{- s_0 +\beta\epsilon(x^*)} \right]^{-1}.
\end{equation}
Here, $\epsilon(x^*)$ is the population-dependent average excitation energy
\begin{equation}
  \label{eq:2-19}
  \epsilon(x^*) = \epsilon_0 - 2x^*\lambda_2 + 2(1-x^*)\lambda_1 .
\end{equation}
Mechanical stability of the ideal glass state requires 
\begin{equation}
  \label{eq:2-19-1}
  \epsilon(x^*) > 0.
\end{equation}

Equation (\ref{eq:2-18}) indicates that the introduction of finite
widths to the two RS delta functions of the two-state model results in
self-consistency in determining the excited-state population which is
governed by the average ground-to-excited energy gap $\epsilon(x)$ [cf.\ 
Eqs.\ (\ref{eq:2-13}) and (\ref{eq:2-18})].  The function $\epsilon(x)$ has a
simple physical meaning. Going from two states with the gap $\epsilon_0$ to a
disordered material with Gaussian distributions of the RS ground and
excited state energies makes states with lower random energies
thermodynamically more probable.  This effectively lowers the energy
of each RS state by ``solvation'' energy $2\lambda_i$ (Stokes shift in
spectroscopic applications). The lowering of the energy of each state
is, however, scaled with the corresponding population and one gets
$2x\lambda_2$ and $2(1-x)\lambda_1$ for lowering the excited and ground states,
respectively. The dependence on ground and excited state populations
makes the excitation energy decrease with increasing temperature from
$\epsilon_0 + 2\lambda_1$ at low temperature to $\epsilon_0 - 2\lambda_2$ at high temperature
(Fig.\ \ref{fig:11}).

\begin{figure}[htbp]
  \centering \includegraphics*[width=7cm]{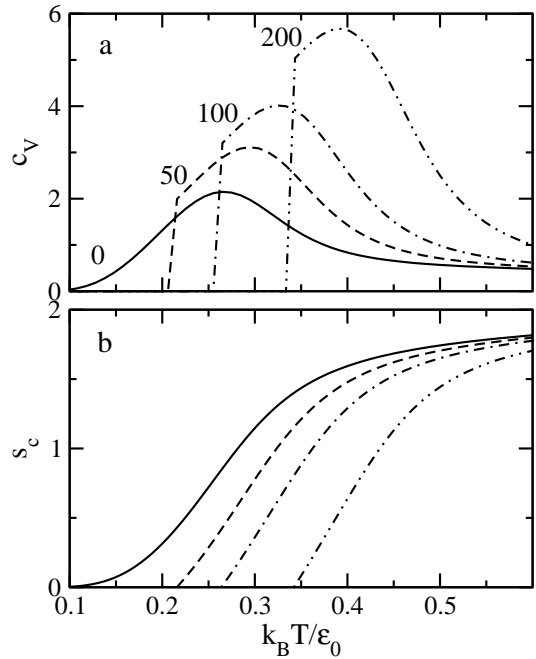}
  \caption{$c_V$ (a) and $s_c$ (b) (in $k_{\text{B}}$ units) vs 
    temperature at $\lambda_1=0$ (solid line), $\lambda_1/k_{\text{B}}=50$ K
    (dashed line), $\lambda_1/k_{\text{B}}=100$ K (dash-dotted line), and
    $\lambda_1/k_{\text{B}}=200$ K (dash-double-dotted line);
    $\lambda_2/k_{\text{B}}=200$ K, $\epsilon_0/k_{\text{B}}=800$ K, $s_0=2.0$,
    $\lambda_1$ and $\lambda_2$ are assumed to be temperature independent. 
    Numbers on plot (a) indicate values of $\lambda_1/k_{\text{B}}$.}
  \label{fig:4}
\end{figure}

The second thermodynamic relation [Eq.\ (\ref{eq:2-16})] allows one to
connect the width parameter $\sigma$ in the CS Gaussian term in Eq.\ 
(\ref{eq:2-10}) and the constant-volume heat capacity per molecule
$c_V(T)$. The width in the Gaussian term can be separated into the
$k_{\text{B}}T$ factor and an energetic coupling parameter related to
the RS coupling constants $\lambda_i$ [Eqs.\ (\ref{eq:2-21-1}) and
(\ref{eq:2-20})]:
\begin{equation}
  \label{eq:2-21}
    \sigma^2 = 2 k_{\text{B}}T \lambda_0(T), 
\end{equation}
The form of the CS Gaussian width in Eq.\ (\ref{eq:2-21}) is dictated
by the classical limit of the fluctuation-dissipation theorem
(FDT).\cite{Landau5} The energy parameter $\lambda_0(T)$ then plays the role
of the trapping energy in theories of random-media
conductivity\cite{Baessler:87} or the solvent reorganization energy in
theories of electron transfer.\cite{Marcus:93} Alternatively, $\lambda_0(T)$
enters the constant volume heat capacity obtained from Eqs.\ 
(\ref{eq:2-10}) and (\ref{eq:2-16})
\begin{equation}
  \label{eq:2-22}
  c_V(T) = 2\beta \lambda_0(T)\left[1 - \alpha^*(\epsilon_0 - 4x^*\lambda_2 +4(1-x^*)\lambda_1) \right]^{-1} ,
\end{equation}
where
\begin{equation}
  \label{eq:2-23}
  \alpha^* = \frac{dx(\phi)}{d\phi}\bigg|_{\phi=\phi^*} .
\end{equation}

The separation of the width $\sigma^2$ in the $k_{\text{B}}T$ factor and
$\lambda_0(T)$ is convenient when the latter is only weakly dependent on
temperature.  This implies that $c_V(T)$ is approximately a hyperbolic
function of temperature $c_V(T)\propto 1/T$ as is empirically documented, at
least for the constant pressure heat
capacity.\cite{Privalko:80,Aba:90} The temperature dependence of $\lambda_0
(T)$ [Eq.\ (\ref{eq:2-21-1})] is determined by the gap $\Delta \lambda = \lambda_2 -
\lambda_1$ in the coupling parameters between the ground and excited states
since $\lambda_0 (T)$ has a sigmoidal form (Fig.\ \ref{fig:3}) with the
change $\Delta \lambda$ from low to high temperatures ($\lambda_i$ are assumed to be
temperature independent in Fig.\ \ref{fig:3}). A qualitatively similar
sigmoidal form often observed in
simulations\cite{Sastry:98,Jung:04,Doliwa:03,Chowdhary:04} is seen for
the basin depth $\phi^*$. The basin depth increases by the amount $\epsilon_0 -
2\Delta\lambda$ with increasing temperature (Fig.\ \ref{fig:3}).

The description of the glass thermodynamics in terms of a distribution
of basin energies involves the projection of the whole configuration
space onto a single collective coordinate. This projection onto the
lower dimension\cite{Feynman:63} introduces a reduced description of
the system thermodynamics reflected by the temperature dependence of
the moments of the collective coordinate, as is well known from e.g.\ 
electron transfer theory.\cite{Marcus:93} In fact, thermodynamics
[Eq.\ (\ref{eq:2-16})] and the classical limit of the FDT [Eq.\ 
(\ref{eq:2-21})] both predict that, for fluctuations caused by
classical motions, $\sigma^2$ should decompose into the $k_{\text{B}}T$
factor and a coupling parameter $\lambda_0(T)$. This is a significant
departure from the simple temperature independent behavior of $\sigma^2$
supposed to date. Such a temperature dependence modifies predictions
of even the simple Gaussian model: no ideal glass transition occurs
for temperatures which leave the expression $s_0(x^*) - \beta\lambda_0(T)$
positive. We also note that the combination of disorder with the FDT
leads to the excitation energy $\epsilon(x^*)$ [Eq.\ (\ref{eq:2-19})]
decreasing with temperature.

The distribution of basin energies is affected by temperature through
the explicit $k_{\text{B}}T$ factor in Eq.\ (\ref{eq:2-21}) and
through a more complex temperature variation of $\lambda_0(T)$.  Figure
\ref{fig:2}c shows $\omega(\phi)$ at different temperatures obtained under the
assumption that $\lambda_i$ defined by Eq.\ (\ref{eq:2-20}) are temperature
independent. The distribution, which is almost Gaussian at high
temperatures, gets skewed from the high-energy wing at lower
temperatures (cf.\ solid and dash-dotted lines in Fig.\ \ref{fig:2}c).
The low-energy wing of the enumeration function is not strongly
affected by temperature, and low-energy wings at different
temperatures can approximately be brought to one master curve by a
vertical shift.  The high-energy wings differ, however, substantially
as temperature changes. Exactly this behavior of the enumeration
function curves at different temperatures is reported for the 80-20 LJ
binary mixture in Fig.\ 4 of Ref.\ \onlinecite{Sciortino:00}.

The enumeration function taken at the average basin depth $\phi^*$ gives
the configurational entropy (in $k_{\text{B}}$ units)
\begin{equation}
  \label{eq:2-25}
       s_c  = s_0(x^*) - \beta\lambda_0 . 
\end{equation}
This equation predicts the existence of ideal glass transition,
$s_c(T_K)=0$, at a finite temperature when $\lambda_1\neq 0$. The Kauzmann
temperature $T_K$ tends to zero when $\lambda_1\to 0$ and $s_c>0$ for any
$0<x^*<1$ at $\lambda_1=0$ (solid line in Fig.\ \ref{fig:4}b). The constant
volume heat capacity can directly be calculated from Eq.\
(\ref{eq:2-22}). Calculations of $c_V(T)$ at constant $\lambda_2$ and
varying $\lambda_1$, both temperature independent, are shown in Fig.\
\ref{fig:4}a. At $\lambda_1\neq 0$ the heat capacity drops to zero at the point
of ideal glass transition at $T=T_K>0$.  There is no ideal glass
transition when $\lambda_1=0$, and the heat capacity passes through a broad
maximum.

\begin{widetext}
\begin{table*}
  \caption{Best-fit parameters for model fluids (rows 1-2) and real liquids 
           (rows 3-6). For real liquids, 
           the fitting parameters are $\lambda_1$, 
           $\lambda_2$, $\epsilon_0$ (K) and $s_0$ are obtained by global fits of 
           Eqs.\ (\ref{eq:2-19}) and (\protect{\ref{eq:3-1}}) 
           to experimental heat capacities and entropies.  }
  \label{tab:1}
  \centering
  \begin{tabular}{lcccccccccc}
\hline
Substance & $z$ & $T_g$ & $T_{\text{fus}}$ & $\Delta s_{\text{fus}}$\footnotemark[1] & 
        $\epsilon_0/k_{\text{B}}$ & 
        $\lambda_1/k_{\text{B}}$ & $\lambda_2/k_{\text{B}}$ & $s_0$\footnotemark[1] & $T_K$\footnotemark[2] 
        & $T_K$\footnotemark[3] \\
\hline
80-20 BLJM\footnotemark[4] 
         & 1 &    &          &       & 69.3 & 8.9 & 26.5 & 0.14 & 34.35\footnotemark[5] & 30.4 \\ 
50-50 BSSM\footnotemark[6]
         & 1 &    &          &       & 78.5 & 2.0 & 22.3 & 1.59    &       &  14.2  \\
Glycerol & 8  & 190 & 291      & 7.55  & 631  & 18 & 32    & 1.27   & 136.7 & 130   \\
Selenium & 1  & 304 & 494.33   & 1.50  & 905  & 14  & 27    & 1.7  & 210.7 & 153 \\
Toluene  & 2  & 117 & 178.15   & 4.48  & 1045  & 36  & 518   & 5.1  & 99.9  & 110\footnotemark[7] \\
$o$-terphenyl & 2 & 246 & 329.4  & 6.28  & 2267 & 45  & 1133  & 5.5  & 204.1 & 214\footnotemark[7] \\
\hline
  \end{tabular}
\footnotetext[1]{In $k_{\text{B}}$ units.}
\footnotetext[2]{Kauzmann temperature (K) from extrapolation of experimental configurational entropy $s_c$.}
\footnotetext[3]{Kauzmann temperature (K) from the condition $s_c(T_K)=0$ obtained from the 2G model.}
\footnotetext[4]{Binary LJ A-B mixture with $\epsilon_{AA}/k_{\text{B}}= 119.8$ K, 
                 $\epsilon_{AB}/ \epsilon_{AA}= 1.5$, $\epsilon_{BB}/ \epsilon_{AA}= 0.5$, 
                 $\sigma_{BB}/ \sigma_{AA}=0.88$, and density $\rho/ \rho_0=1.2$,
                 $\rho_0=2.53×10^{28}$ m$^3$.}
\footnotetext[5]{Calculated from $s_c(T_K)=0$ in Ref.\ \onlinecite{Sastry:01}. $T_K$ obtained from the
potential energy landscape method is 34.7 K, temperature of vanishing diffusivity from
the Vogel-Fulcher-Tammann plot is $T_0=35.97$ K. }
\footnotetext[6]{Binary soft sphere mixture. 
 The reduced energy parameters from the fit are multiplied with $\epsilon_{AA}/k_{\text{B}}= 119.8$ K 
                 for consistency with the 80-20 BLJM.}
\footnotetext[7]{Determined as the temperature of the first-order phase transition at which the entropy
                 discontinuously drops to zero. }
\end{table*}
\end{widetext}

\section{Liquid-Liquid(Glass) First Order Phase Transition}
\label{sec:6}
Depending on the value of the excitation entropy $s_0$ Eq.\
(\ref{eq:2-18}) may have one, two, or three solutions. Only one
solution exists for low $s_0$ since the condition of mechanical
stability requires that the effective excitation energy $\epsilon(x^*)$
remains positive at all excited state populations [Eq.\
(\ref{eq:2-19-1})]. Increasing the entropy of excitation allows
one to reach the condition of vanishing Gibbs energy of excitation
\begin{equation}
  \label{eq:6-1}
  g(x^*) = \epsilon(x^*) - k_{\text{B}} T s_0
\end{equation}
in the range $0\leq x^* \leq 1$. At the temperature 
\begin{equation}
  \label{eq:6-2}
  T_{LL} =(\epsilon_0 - \Delta\lambda )/s_0
\end{equation}
the excitation Gibbs energy vanishes at $x^*=1/2$, $g(1/2)=0$. This is
the temperature of the equilibrium liquid-liquid, first-order phase
transition between the low-temperature liquid with low concentration
of excitations and the high-temperature liquid with high concentration
of excitations.\cite{Sastry:03,Tanaka:04} This transition becomes a
liquid-glass transition when the low-temperature phase has its
viscosity below the glass transition limit or when the transition
occurs directly to the ideal glass state.

\begin{figure}[htbp]
  \centering \includegraphics*[width=6cm]{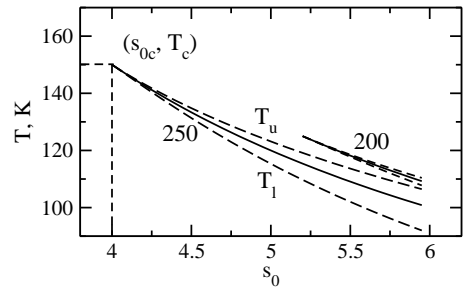}
  \caption{Equilibrium first-order transition temperature (solid
  lines) and the lower, $T_l$, and upper, $T_u$, spinodal temperatures
  (dashed lines) vs the excitation entropy $s_0$. The calculations are
  carried out for $\epsilon_0/k_{\text{B}} = 800$ K, $\lambda_1=50$ K, and values
  of $\lambda_2/k_{\text{B}}$ indicated on the plot; $(s_{0c},T_c)$ indicates
  the critical point. } \label{fig:14}
\end{figure}

The first order transition is possible for temperatures below the
critical temperature $T_c$ and excitation entropies above the critical
value $s_{0c}$:
\begin{equation}
  \label{eq:6-3}
  T<T_c,\quad s_0>s_{0c} .
\end{equation}
The critical parameters are
\begin{equation}
  \label{eq:6-4}
  \begin{split}
  k_{\text{B}}T_c &= \bar\lambda=(\lambda_1+\lambda_2)/2,\\ 
  s_{0c}       &= \frac{\epsilon_0 - \Delta \lambda}{\bar\lambda} .
  \end{split} 
\end{equation}
The critical temperature $T_c$ increases and the critical entropy
$s_{0c}$ decreases with enhanced disorder of the excited state (Fig.\ 
\ref{fig:14}). When $T<T_c$ and $s_0 >s_{0c}$, the equilibrium
transition temperature, $T_{LL}$, is flanked by lower, $T_l$, and
upper, $T_u$, spinodal temperatures at which only two solutions of
Eq.\ (\ref{eq:2-18}) are possible: $T_l\leq T_{LL}\leq T_u$ (Fig.\ 
\ref{fig:14}).

\begin{figure}[htbp]

 \centering \includegraphics*[width=6cm]{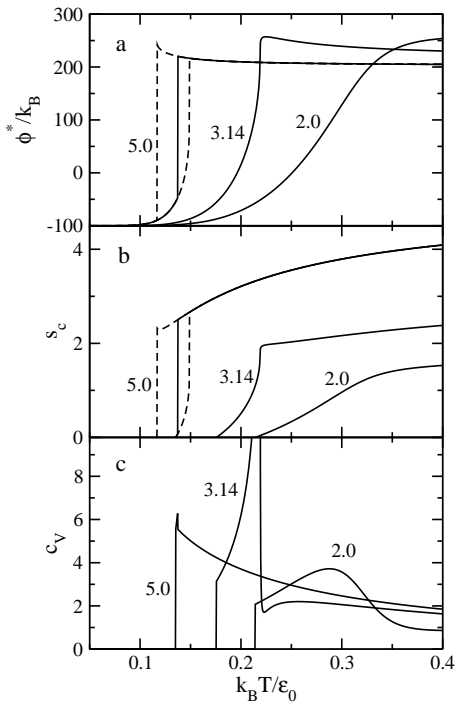}
\caption{Basin energy (a), configuration entropy (b), and constant-volume
         heat capacity (c) at the values of $s_0$ indicated on the plot; 
         $\lambda_1/k_{\text{B}}=50$ K, $\lambda_2/k_{\text{B}}=300$ K, $\epsilon_0/k_{\text{B}}=800$ K. The values
         $s_0=3.14$ and $k_{\text{B}}T/ \epsilon_0 = 0.22$ correspond to the critical point 
         at which lambda singularity for the heat capacity is seen. The dashed
         lines indicate metastable states between the lower and upper spinodal
         temperatures. }
\label{fig:15}
\end{figure}

Figure \ref{fig:15} illustrates the change in the temperature
variation of the thermodynamic parameters with increasing the
excitation entropy $s_0$.  At $s_0<s_{0c}$, the entropy and basin
energy are both continuous functions of temperature. The heat capacity
passes through a broad maximum characteristic of the two-state model
and drops to zero at a Kauzmann temperature $T_K>0$ when $\lambda_1\neq 0$. At
the critical excitation entropy $s_0=s_{0c}$, the entropy and energy
both pass through an inflection point reflected by a lambda
singularity in the constant volume heat capacity. Finally, the
liquid-liquid (glass) phase transition occurs above $s_{0c}$. The
entropy drop at the transition temperature increases with increasing
$s_0$ to the point where the entropy drops to zero. This transition, as
well as all transitions with higher entropy $s_0$, occur directly from
the supercooled liquid to the ideal glass state.

\section{Comparison to simulations}
\label{sec:3}
The temperature variation of the energy landscape is critically
affected by the approximately linear temperature dependence of the
width parameter $\sigma$ [Eq.\ (\ref{eq:2-21})] which contributes to the overall temperature
variation of the distribution of basin energies
\begin{equation}
  \label{eq:3-1}
  P(\phi)\propto e^{\omega(\phi) - \beta\phi} .
\end{equation}
This generally non-Gaussian distribution is often approximated by a Gaussian function:
\begin{equation}
  \label{eq:3-2}
  P(\phi) \propto \exp \left( - \frac{(\phi-\phi^*)^2}{2\Gamma^2}\right)
\end{equation}
with the empirical Gaussian width $\Gamma$. Computer simulations of
$P(\phi)$ and the constant volume heat capacity $c_V$ at varying
temperature may provide insights into the temperature dependence of
$\sigma$.  Approximately Gaussian distribution of basin energies has
been found in simulations of binary
LJ\cite{Speedy:88,Buchner:99,Sciortino:00,Sastry:01} and hard sphere
fluids.\cite{Speedy:88} Although an increase of the width with
temperature is often seen in simulations,\cite{Sastry:98,Sciortino:00}
numerical data are very limited.

\begin{figure}
  \centering \includegraphics*[width=8.5cm]{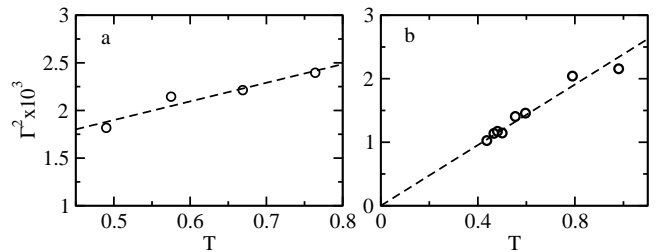}
  \caption{Variance of the distribution of metabasins in the A-B 
    binary LJ fluid determined from simulations by Denny \textit{et
      al}\cite{Denny:03,com:2} (a) and by Doliwa and
    Heuer\cite{Doliwa:03} (b).  Units of energy and temperature are
    defined by the LJ energy of the A-A interaction potential. }
  \label{fig:5}
\end{figure}

Temperature-dependent width $\Gamma(T)$ can be found in simulations of the
well-known 80-20 LJ mixture\cite{Weber:85,Kob:95} by B\"uchner and
Heuer.\cite{Buchner:99,Heuer:00} Also, recent extensive simulations by
Denny \textit{et al.}\cite{Denny:03} and by Doliwa and
Heuer\cite{Doliwa:03} give the variance of energies of metabasins
(basins separated by small minima which do not require activated hops
at a given temperature\cite{Stillinger:95,Doliwa:03}) for the same
system at different temperatures.  The width of metabasin distribution
$\Gamma(T)$ extracted\cite{com:2} from the simulations by Denny \textit{et
  al.}\cite{Denny:03} is approximately linear in $T$ (Fig.\ 
\ref{fig:5}a). An even steeper $\Gamma(T)$ is found in simulations by
Doliwa and Heuer\cite{Doliwa:03} of the same system of a smaller size
(Fig.\ \ref{fig:5}b).

\begin{figure}[htbp]
  \centering \includegraphics*[width=8.5cm]{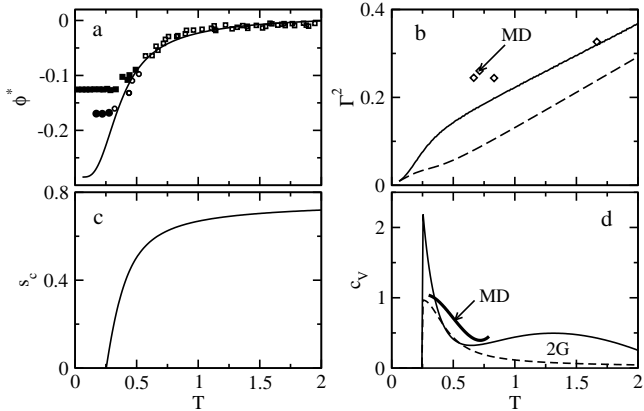}
  \caption{Inherent structure energy (a), the distribution width (b), 
    configurational entropy (c), and constant volume heat capacity (d)
    for the 80-20 LJ binary mixture.\cite{Kob:95,Sastry:98,Sastry:01}
    The solid lines refer to calculations with the present model with
    the parameters obtained from the fit of inherent structure
    energies from Ref.\ \onlinecite{Sastry:98} combining data for the
    cooling rate 2.7 $× 10^{-4}$ (squares) and 3.33$× 10^{-6}$
    (circles).  The closed points in (a) indicate the points which
    were excluded from the fit since they represent the loss of
    system's ergodicity due to finite cooling rate.  In (b), the solid
    line refers to $\Gamma^2$, the dashed line refers to $\sigma^2$, and ``MD''
    refers to simulation results for $\Gamma^2$ by B\"uchner and
    Heuer.\cite{Buchner:99,Heuer:00} In (d), the dashed line refers to
    the constant $P$ heat capacity calculated under assumption that
    $\lambda_i$ are temperature independent. The bold solid line marked
    ``MD'' is obtained by numerical differentiation of the
    configurational entropy from MD simulations by Sciortino
    \textit{et al.}\cite{SciortinoPRL:99} and ``2G'' marks the present
    model.  Units of energy and temperature are defined by the LJ
    energy of the A-A interaction potential. The parameters of the fit
    are listed in Table \ref{tab:1}. }
  \label{fig:10}
\end{figure}

The 2G model is based on four parameters: $\epsilon_0$, $\lambda_1$, $\lambda_2$, and
$s_0$.  The model is tested on its ability to reproduce several
thermodynamic observables for a single set of parameters. We first
apply the 2G model to simulations of model fluids and then, in Sec.\ 
\ref{sec:4}, apply it to real liquids.  For comparison to computer
experiment we fit the average basin energy from 2G model to combined
ergodic parts of two cooling runs reported in Ref.\ 
\onlinecite{Sastry:98} for the 80-20 binary LJ mixture (Fig.\ 
\ref{fig:10}a, Table \ref{tab:1}).  The fitting parameters obtained
for the basin depth are used to calculate the configurational entropy
[Eq.\ (\ref{eq:2-25})] which goes to zero (Fig.\ \ref{fig:10}c) at the
Kauzmann temperature very close to that obtained from simulations of
$s_c$ itself and to $T_0$ from diffusivity extrapolated to zero
through the Vogel-Fulcher-Tammann plot (cf.\ columns 10 and 11 in Table
\ref{tab:1}).

Since the distribution $P(\phi)$ is generally non-Gaussian, the empirical
width $\Gamma$ was obtained from the half-intensity width of $P(\phi)$.  $\Gamma^2$
(solid line in Fig.\ \ref{fig:10}b) is larger than $\sigma^2$ (dashed line
in Fig.\ \ref{fig:10}) since the former reflects the overall width
arising from the ideal mixture entropy and the CS Gaussian term in
Eq.\ (\ref{eq:2-10}) combined.  The empirical width $\Gamma^2(T)$ rises
with temperature in accord with the prediction of the FDT and the
results of simulations.  The magnitude of $\Gamma^2$ for the distribution
of basins is significantly larger than the one for the distribution of
metabasins (cf.\ Fig.\ \ref{fig:5} and Fig.\ \ref{fig:10}b). However,
the variance of basin energies from MD simulations by B\"uchner and
Heuer\cite{Buchner:99,Heuer:00} (marked ``MD'' in Fig.\ \ref{fig:10}b)
is in reasonable agreement with the 2G model.  Finally, the constant
volume heat capacity shows a steep rise on approach to the ideal glass
transition, dropping to zero at $T_K$.  This form of the heat capacity
is supported by simulations of Sciortino \textit{et
  al}.\cite{SciortinoPRL:99} as shown in Fig.\ \ref{fig:10}d. We note
that the 80-20 system was originally parameterized to represent the
metallic Ni-P alloy\cite{Weber:85} and that metallic glassformers
typically show very sharp excess heat capacity functions relative to
molecular and ionic glassformers.\cite{Angell:95,Busch:00}

\begin{figure}[htbp]
  \centering
  \includegraphics*[width=8.5cm]{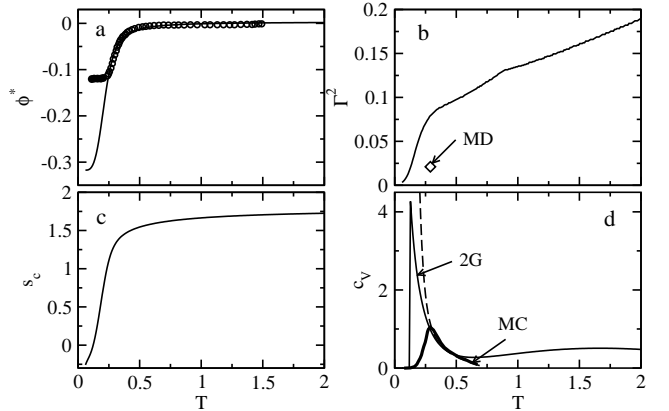}
  \caption{Same as in Fig.\ \ref{fig:10} for the 50-50 BSSM studied by 
    Yan \textit{et al}.\cite{Yan:04}. The parameters of the fit are
    listed in Table \ref{tab:1}. In (b), ``MD'' marks the Gaussian
    width from parallel tempering MD simulations reported in Refs.\ 
    \onlinecite{Yu:02} and \onlinecite{Yu:04}.  In (d), ``2G'' marks
    the present two-Gaussian model, the bold line marked ``MC'' refers
    to the results of Monte Carlo simulations from Ref.\ 
    \onlinecite{Yan:04}.  The dashed line shows the fit of the $c_V$
    simulation data\cite{Yu:04} extrapolated below the temperature of
    heat capacity drop. The functional form of $c_V(T)$ (dashed line)
    is from Ref.\ \onlinecite{Yu:04}: $c_V(T) = A_1T^{B_1} +
    A_2T^{B_2}$, $A_1=2.845$, $B_1=-0.209$, $A_2=3.477×10^{-4}$, $B_2
    = -5.804$. }
  \label{fig:12}
\end{figure}

A similar fit of the 2G model to the average basin depth of the 50:50
soft sphere (SS) mixture reported by Yan \textit{et al}.\cite{Yan:04}
is shown in Fig.\ \ref{fig:12}. The parameters of the fit are used to
calculate the distribution width $\Gamma$ (Eq.\ (\ref{eq:3-2}), Fig.\ 
\ref{fig:12}b), the configurational entropy (Eq.\ (\ref{eq:2-25}), Fig.\ \ref{fig:12}c), and
the constant volume heat capacity (Eq.\ (\ref{eq:2-22}), Fig.\ 
\ref{fig:12}d).  The basin energy width calculated from 2G model ($\Gamma/
\epsilon_{AA} = 0.08$) is higher than the one observed in
simulations\cite{Yu:02,Yu:04} ($\Gamma/\epsilon_{AA} = 0.02$, marked ``MD'' in Fig.\ 
\ref{fig:12}b).

Despite the use of parallel tempering MD in Refs.\ \onlinecite{Yu:02}
and \onlinecite{Yu:04}, the drop of the heat capacity at the peak
temperature $T_p\simeq 36.5$ K ($\epsilon_{AA}/k_{\text{B}} = 119.8$ K) seen in
the simulations is due to insufficient sampling of the phase
space.\cite{Yu:02,Yu:04} The parameters obtained from the fit of the
average basin depth (Table \ref{tab:1}) give a reasonable description
of the simulated $c_V(T)$ up to $T_p$ followed by a much stronger rise of
$c_V(T)$ which drops to zero at $T_K\simeq 14 $ K. This latter temperature is
close to the point of vanishing configurational heat capacity in the
simulations. Note that a peak of $c_V(T)$ higher that the one reported in
Refs.\ \onlinecite{Yan:04}, \onlinecite{Yu:02}, and \onlinecite{Yu:04}
was obtained in MC simulations of analogous binary SS mixture by
Grigera and Parisi\cite{Grigera:01} within a simulation protocol
outperforming parallel tempering.  This implies that, once sampling is
improved, $c_V(T)$ continues to grow beyond the drop at $T_p$.  Also note
that extrapolation of $c_V(T)$ from the fit of simulation data by Yu
and Carruzzo\cite{Yu:04} to lower temperatures goes even steeper
(dashed line in Fig.\ \ref{fig:12}d) than $c_V(T)$ from 2G model.

\section{Experimental configurational entropies and the Kauzmann temperature}
\label{sec:4}
The configurational heat capacity of a liquid is normally taken as the
difference between the liquid and crystal entropies reported for
constant pressure, although it is known that in many cases a part of
the entropy of fusion is due to an increase in the vibrational entropy
(arising from increases in the low frequency vibrational density of
states in the liquid inherent
structures\cite{Goldstein:76,Angell:04,Chowdhary:04}).  The constant $P$
configurational heat capacity can be calculated from the
configurational entropy in Eq.\ (\ref{eq:2-25})
\begin{equation}
  \label{eq:4-1}
  c_P = T\left( \frac{\partial s_c}{\partial T}\right)_P. 
\end{equation}
The unknown parameter in this calculation is the temperature
dependence of the model parameters $\epsilon_0$, $\lambda_i$, and $s_0$ at constant
$P$. Spectroscopic measurements at constant $P$ give ``solvation
energies'' $\lambda_i$ through spectral Stokes shifts\cite{Ranko:00} which
are weakly temperature dependent.\cite{DMjpcb:99} The Stokes shift
relates to the coupling of a localized state to a thermal Gaussian
bath.  Since the defect excitations considered here may be more or
less delocalized, it is currently unclear if the assumption $(\partial \lambda_i/ \partial
T)_P=0$ is warranted.  An alternative scenario might include no
temperature dependence of the ideal glass distribution $\sigma_1$
corresponding to quenched disorder (i.e., $\lambda_1\propto 1/T$) and a standard
dependence on temperature of $\sigma_2$ (i.e., $\lambda_2=Const$). It turns out
that, when the 2G model is applied to fit the experimental excess heat
capacities of the liquid over the crystal, $\Delta c_P = c_{P,\text{liq}} -
c_{P,\text{cryst}}$, the results are fairly insensitive to the
assumptions made regarding the temperature dependence of $\lambda_1$ once
the condition $\lambda_2(T)=Const$ is adopted. The fit of experimental results
is thus done with temperature-independent $\epsilon_0$, $\lambda_i$, and $s_0$.

\begin{figure}[htbp]
  \centering \includegraphics*[width=8cm]{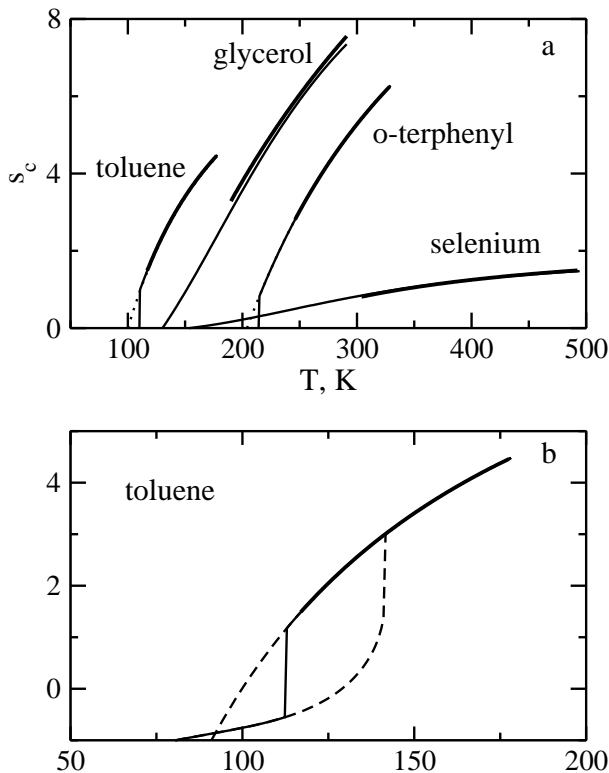}
  \caption{(a): Configurational entropy $s_c$ vs temperature for liquids listed in Table \ref{tab:1}.
    Thick solid lines are experimental results and the thin solid
    lines are fits to the 2G model.  Experimental data stop at the
    glass transition temperature $T_g$. Dotted lines show the
    extrapolation of experimental entropies to the zero entropy line.
    The fitting parameters are listed in Table \ref{tab:1}. (b):
    Temperature dependence of the entropy of toluene (scaled up). The
    dashed lines indicate the entropies of metastable states
    terminated at the lower and upper spinodal temperatures.  }
  \label{fig:6}
\end{figure}

The fitting procedure involves simultaneous fit of Eqs.\
(\ref{eq:2-25}) and (\ref{eq:4-1}) with four fitting parameters,
($\lambda_1$, $\lambda_2$, $\epsilon_0$, and $s_0$) to experimental heat capacities $\Delta
c_P$ and experimental configurational entropies.\cite{Moynihan:00} The
range of energy parameters is restricted by the condition of
mechanical stability of the ideal glass state [Eq.\
(\ref{eq:2-19-1})].  The configurational entropy at constant $P$ can
be determined experimentally from the entropy of fusion $\Delta
s_{\text{fus}}$ and $\Delta c_P(T)$ (both in $k_{\text{B}}$ units)
\begin{equation}
  \label{eq:4-2}
  s_c(T) = \Delta s_{\text{fus}} + \int_{T_{\text{fus}}}^{T} (\Delta c_P(T')/T')dT' .
\end{equation}

The 2G model outlined in Sec.\ \ref{sec:2} assumes that each molecule
represents one excitable unit. While this is true for atomic glasses
like selenium, for more complex compounds one needs to introduce the
number $z$ of independently excitable (i.e. rearrangeable) states per
molecule or formula unit.\cite{Moynihan:00} The parameter $z$ is taken
from Takeda \textit{et al}.\cite{Takeda:99} (Table \ref{tab:1}) and is
used to multiply the heat capacity in Eq.\ (\ref{eq:4-1}) in fitting
the experimental data. The results of the fit for four glassformers
are listed in Table \ref{tab:1}.

\begin{figure}[htbp]
  \centering
  \includegraphics*[width=8cm]{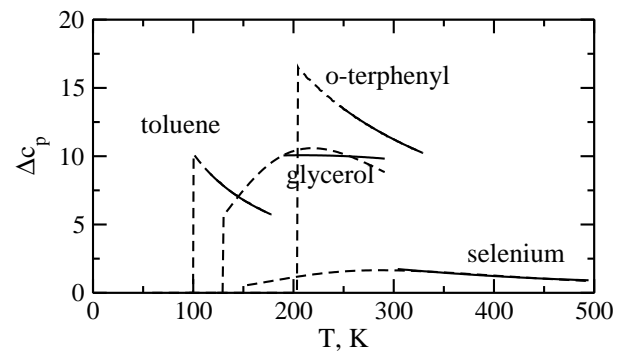}
  \caption{Comparison of the experimental values of $\Delta c_P$ (solid lines) with 
    $\Delta c_P$ calculated from the 2G model (dashed lines) by 
    simultaneous fit to the experimental $s_c$ and $\Delta c_P$ data above $T_g$. }
  \label{fig:7}
\end{figure}

The examination of Table \ref{tab:1} shows that $\lambda_1>0$ for all fluid
studied, indicating that $T_K>0$.  The distribution of excited states
is much narrower in the case of covalent and hydrogen-bonded liquids
(selenium and glycerol) compared to molecular liquids (toluene and
$o$-terphenyl). In contrast, the RS distribution of the ideal glass is
almost invariant among different glassformers.  When $\lambda_2\gg\lambda_1$ the
configurational entropy $s_c(T)$ gains a bend close to the Kauzmann
point (Fig.\ \ref{fig:6}) resulting in the actual $T_K$ from
$s_c(T_K)=0$ smaller than the corresponding value from extrapolation
of experimental entropies (e.g., selenium in Table \ref{tab:1}).

\begin{figure}[htbp]
  \centering
  \includegraphics*[width=8cm]{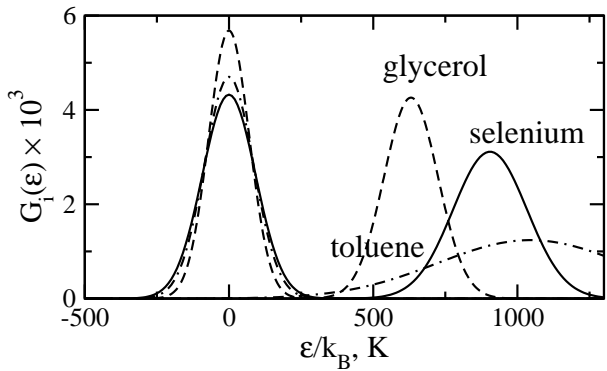}
  \caption{Real-space distribution of energies for selenium (solid lines), glycerol (dashed lines),
    and toluene (dash-dotted lines) at their corresponding Kauzmann
    temperatures. The parameters $\epsilon_0$ and $\lambda_i$ used in the
    calculations are taken from Table \ref{tab:1}.}
  \label{fig:8}
\end{figure}

The most interesting result of our analysis is the low-temperature
behavior of fragile molecular glasses (toluene and $o$-terphenyl).
These substances are characterized by high disorder of the excited
state ($\lambda_2\gg\lambda_1$, Fig.\ \ref{fig:8}) and, in addition, high entropy of
excitation (Table \ref{tab:1}). It also turns out that $s_0$ from the
fit is higher than the critical excitation entropy which is close to
2.0 for both liquids. The fact that $s_0$ is more than twice higher
than $s_{0c}$ ensures low first-order transition temperature, well
below the critical temperature (543 K for toluene and 1156 K for
$o$-terphenyl). The first-order transition temperature in fact falls
in the unobservable range between $T_K$ and $T_g$ where the entropy
discontinuously drops to zero producing a similar drop in the heat
capacity (Figs.\ \ref{fig:6} and \ref{fig:7}). It may be therefore
suggested that fragile liquids resolve the Kauzmann paradox by a first
order liquid-glass transition. We note that both for toluene and
$o$-terphenyl the lower spinodal temperature is below the point when
metastable entropy crosses the zero entropy line while the upper
spinodal temperature is above $T_g$ for toluene and almost coincides
with $T_g$ for $o$-terphenyl. Since the first order transition is
below $T_g$, the equilibrium passage along the solid line in Fig.\ 
\ref{fig:6}b is unlikely thus suggesting hysteresis of the heat
capacity between the cooling and heating runs.

The value $s_0$, which remains an empirical parameter of the 2G model,
can be compared to the entropy cost of creating a density wave in 
density-functional theories of aperiodic structures.\cite{Dasgupta:99,Xia:00} 
The average entropy of the ``entropy droplet'' in Wolynes's mosaic model is
\begin{equation}
  \label{eq:4-3}
  s_0 = \frac{3}{2} \ln\left( \alpha r_0^2 / \pi\right) - \frac{5}{2} ,
\end{equation}
where $\alpha$ represents the rms displacement from the lattice site and
$r_0$ is the mean lattice spacing. Invoking the Lindemann
ratio\cite{Xia:00} $\alpha^{1/2}r_0 = 10$, one gets $s_0=2.7$, which falls
in between entropies for strong and fragile liquids in Table
\ref{tab:1}.

Caution is needed in these interpretations since the model is in the
early stages of evaluation and there are four parameters even for
simply constituted glasses ($z = 1$). One of these parameters may be
disposable. It is apparent from Table \ref{tab:1} and Fig.\ 
\ref{fig:8}, that the ground RS Gaussian width best fitting
the various data, while non-zero (as expected for a non-crystalline
ground state), is small relative to the excited state Gaussian (except
for the stronger liquids) and not varying much between the different
systems. It could probably be given a fixed value, reducing the
disposable parameters to 3 for simple glasses and 4 for flexible
molecule glasses, where the 4th parameter can be fixed from molecular
considerations.\cite{Takeda:99}

\section{Concluding Remarks}
We have shown that by introducing a realistic form for defect-like
excitations in glasses, the basic ``excitations'' model of the glass
transition can be developed in a form that bridges the gap between
previous over-simple models and the random energy model of
Derrida.\cite{Derrida:80} In other words, we have provided a physical
basis for the previously empirical ``logarithmically modified
Gaussian'' model of Debenedetti \textit{et al}.\cite{Debenedetti:03}
The model is fundamentally non-Gaussian in configuration space.  It
recognizes the role of fluctuations within the FDT in making the
landscape temperature-dependent as evidenced by the behavior of
(meta)basin energy variances from molecular dynamics
simulations.\cite{Buchner:99,Heuer:00,Denny:03,Doliwa:03,Yu:02,Yu:04}

The model predicts a possibility of first-order liquid-liquid (glass)
transition when the entropy of excitations exceeds its critical value
and the temperature falls below the critical point. For fragile
liquids characterized by a broad distribution of excitation energies
and high entropy change per excitation the transition temperature is
low. While most known liquid-liquid transitions for strong liquids are
at high temperatures,\cite{Sastry:03} the observation of such a
transition for fragile supercooled triphenyl phosphite\cite{Tanaka:04}
supports this trend.  The fit of the model to experimental entropies
and heat capacities of fragile toluene and $o$-terphenyl results in
the first-order liquid/ideal glass transition between $T_g$ and the
experimental Kauzmann temperature. It seems therefore reasonable to
suggest that fragile liquids release the excess entropy by a
first-order transition to the glassy state.

The present model belongs to a class of mean-field two-state models in
which the average excitation energy drops linearly with the increase
in the population of the excited state [Eqs.\ (\ref{eq:2-18}) and
(\ref{eq:2-19})]. Negative excitation energies are prohibited by the
condition of mechanical stability, and crossing the zero point of the
excitation Gibbs energy is driven by the excitation entropy the
magnitude of which is correlated with glass
fragility.\cite{AngellRichards:99} Another physical realization of
this model is the coupling between molecular excited states through
long-range interactions. In case of optical excitations of molecules
coupled by long-range dipolar forces the change in the excitation
energy is realized through the reaction field proportional to the
number of excited molecules. A mean-field description, mathematically
equivalent to the present 2G model, then results in transition to
excitonic condensate in molecules coupled through their transition
dipoles\cite{Logan:87} or to a non-polar/paraelectric phase transition
in dipolar two-state fluids.\cite{DMjcp3:05} In the present model,
disorder is responsible for the trapping energy playing the role of
the reaction field in excitonic condensate models.

\begin{acknowledgments}
  The authors are grateful to Srikanth Sastry and Francesco Sciortino
  for enlightening discussions and also to Juan de Pablo and Pablo
  Debenedetti for helpful comments related to their own work in this
  area. This work was supported by the NSF through the grants
  CHE-0304694 (D.\ V.\ M.) and DMR0082535 (C.\ A.\ A.).
\end{acknowledgments}

\bibliographystyle{apsrev}
\bibliography{/home/dmitry/p/bib/chem_abbr,/home/dmitry/p/bib/photosynth,/home/dmitry/p/bib/liquids,/home/dmitry/p/bib/glass,/home/dmitry/p/bib/et,/home/dmitry/p/bib/dm,/home/dmitry/p/bib/dynamics,/home/dmitry/p/bib/ferro}

\begin{thebibliography}{92}
\expandafter\ifx\csname natexlab\endcsname\relax\def\natexlab#1{#1}\fi
\expandafter\ifx\csname bibnamefont\endcsname\relax
  \def\bibnamefont#1{#1}\fi
\expandafter\ifx\csname bibfnamefont\endcsname\relax
  \def\bibfnamefont#1{#1}\fi
\expandafter\ifx\csname citenamefont\endcsname\relax
  \def\citenamefont#1{#1}\fi
\expandafter\ifx\csname url\endcsname\relax
  \def\url#1{\texttt{#1}}\fi
\expandafter\ifx\csname urlprefix\endcsname\relax\def\urlprefix{URL }\fi
\providecommand{\bibinfo}[2]{#2}
\providecommand{\eprint}[2][]{\url{#2}}

\bibitem[{\citenamefont{Williams et~al.}(1955)\citenamefont{Williams, Landel,
  and Ferry}}]{Williams:55}
\bibinfo{author}{\bibfnamefont{M.~L.} \bibnamefont{Williams}},
  \bibinfo{author}{\bibfnamefont{R.~F.} \bibnamefont{Landel}},
  \bibnamefont{and} \bibinfo{author}{\bibfnamefont{J.~D.} \bibnamefont{Ferry}},
  \bibinfo{journal}{J. Am. Chem. Soc.} \textbf{\bibinfo{volume}{77}},
  \bibinfo{pages}{3701} (\bibinfo{year}{1955}).

\bibitem[{\citenamefont{Cohen and Turnbull}(1959)}]{Cohen:59}
\bibinfo{author}{\bibfnamefont{M.~H.} \bibnamefont{Cohen}} \bibnamefont{and}
  \bibinfo{author}{\bibfnamefont{D.}~\bibnamefont{Turnbull}},
  \bibinfo{journal}{J. Chem. Phys.} \textbf{\bibinfo{volume}{31}},
  \bibinfo{pages}{1164} (\bibinfo{year}{1959}).

\bibitem[{\citenamefont{Turnbull and Cohen}(1961)}]{Turnbull:61}
\bibinfo{author}{\bibfnamefont{D.}~\bibnamefont{Turnbull}} \bibnamefont{and}
  \bibinfo{author}{\bibfnamefont{M.~H.} \bibnamefont{Cohen}},
  \bibinfo{journal}{J. Chem. Phys.} \textbf{\bibinfo{volume}{34}},
  \bibinfo{pages}{120} (\bibinfo{year}{1961}).

\bibitem[{\citenamefont{Adam and Gibbs}(1965)}]{Adam:65}
\bibinfo{author}{\bibfnamefont{G.}~\bibnamefont{Adam}} \bibnamefont{and}
  \bibinfo{author}{\bibfnamefont{J.~H.} \bibnamefont{Gibbs}},
  \bibinfo{journal}{J. Chem. Phys.} \textbf{\bibinfo{volume}{43}},
  \bibinfo{pages}{139} (\bibinfo{year}{1965}).

\bibitem[{\citenamefont{Cohen and Grest}(1979)}]{Cohen:79}
\bibinfo{author}{\bibfnamefont{M.~H.} \bibnamefont{Cohen}} \bibnamefont{and}
  \bibinfo{author}{\bibfnamefont{G.}~\bibnamefont{Grest}},
  \bibinfo{journal}{Phys. Rev. B} \textbf{\bibinfo{volume}{20}},
  \bibinfo{pages}{1077} (\bibinfo{year}{1979}).

\bibitem[{\citenamefont{Cohen and Grest}(1981)}]{Cohen:81}
\bibinfo{author}{\bibfnamefont{M.~H.} \bibnamefont{Cohen}} \bibnamefont{and}
  \bibinfo{author}{\bibfnamefont{G.}~\bibnamefont{Grest}},
  \bibinfo{journal}{Adv. Chem. Phys.} \textbf{\bibinfo{volume}{48}},
  \bibinfo{pages}{370} (\bibinfo{year}{1981}).

\bibitem[{\citenamefont{Frederickson and Andersen}(1984)}]{Frederickson:84}
\bibinfo{author}{\bibfnamefont{G.~H.} \bibnamefont{Frederickson}}
  \bibnamefont{and} \bibinfo{author}{\bibfnamefont{H.~C.}
  \bibnamefont{Andersen}}, \bibinfo{journal}{Phys. Rev. Lett.}
  \textbf{\bibinfo{volume}{53}}, \bibinfo{pages}{1244} (\bibinfo{year}{1984}).

\bibitem[{\citenamefont{G{\"o}tze}(1991)}]{Goetze:91}
\bibinfo{author}{\bibfnamefont{W.}~\bibnamefont{G{\"o}tze}}, in
  \emph{\bibinfo{booktitle}{Liquids, freezing and glass transition}}, edited by
  \bibinfo{editor}{\bibfnamefont{J.~P.} \bibnamefont{Hansen}},
  \bibinfo{editor}{\bibfnamefont{D.}~\bibnamefont{Levesque}}, \bibnamefont{and}
  \bibinfo{editor}{\bibfnamefont{J.}~\bibnamefont{Zinn-Justin}}
  (\bibinfo{publisher}{Elsevier}, \bibinfo{address}{Amsterdam},
  \bibinfo{year}{1991}), vol.~\bibinfo{volume}{1}, p. \bibinfo{pages}{287}.

\bibitem[{\citenamefont{Ngai}(1993)}]{Ngai:93}
\bibinfo{author}{\bibfnamefont{K.~L.} \bibnamefont{Ngai}}, \bibinfo{journal}{J.
  Chem. Phys.} \textbf{\bibinfo{volume}{98}}, \bibinfo{pages}{6424}
  (\bibinfo{year}{1993}).

\bibitem[{\citenamefont{Pitts et~al.}(2000)\citenamefont{Pitts, Young, and
  Andersen}}]{Pitts:00}
\bibinfo{author}{\bibfnamefont{S.~J.} \bibnamefont{Pitts}},
  \bibinfo{author}{\bibfnamefont{T.}~\bibnamefont{Young}}, \bibnamefont{and}
  \bibinfo{author}{\bibfnamefont{H.~C.} \bibnamefont{Andersen}},
  \bibinfo{journal}{J. Chem. Phys.} \textbf{\bibinfo{volume}{113}},
  \bibinfo{pages}{8671} (\bibinfo{year}{2000}).

\bibitem[{\citenamefont{Perera and Harrowell}(1999)}]{Perera:99}
\bibinfo{author}{\bibfnamefont{D.~N.} \bibnamefont{Perera}} \bibnamefont{and}
  \bibinfo{author}{\bibfnamefont{P.}~\bibnamefont{Harrowell}},
  \bibinfo{journal}{J. Chem. Phys.} \textbf{\bibinfo{volume}{111}},
  \bibinfo{pages}{5441} (\bibinfo{year}{1999}).

\bibitem[{\citenamefont{Jung et~al.}(2004)\citenamefont{Jung, Garrahan, and
  Chandler}}]{Jung:04}
\bibinfo{author}{\bibfnamefont{Y.~J.} \bibnamefont{Jung}},
  \bibinfo{author}{\bibfnamefont{J.~P.} \bibnamefont{Garrahan}},
  \bibnamefont{and} \bibinfo{author}{\bibfnamefont{D.}~\bibnamefont{Chandler}},
  \bibinfo{journal}{Phys. Rev. E} \textbf{\bibinfo{volume}{69}},
  \bibinfo{pages}{061205} (\bibinfo{year}{2004}).

\bibitem[{\citenamefont{Cang et~al.}(2003)\citenamefont{Cang, Li, and
  et~al.}}]{Cang:03}
\bibinfo{author}{\bibfnamefont{H.}~\bibnamefont{Cang}},
  \bibinfo{author}{\bibfnamefont{J.}~\bibnamefont{Li}}, \bibnamefont{and}
  \bibinfo{author}{\bibfnamefont{V.~N.~N.} \bibnamefont{et~al.}},
  \bibinfo{journal}{J. Chem. Phys.} \textbf{\bibinfo{volume}{118}},
  \bibinfo{pages}{9303} (\bibinfo{year}{2003}).

\bibitem[{\citenamefont{Glotzer and Donati}(1999)}]{Glotzer:99}
\bibinfo{author}{\bibfnamefont{S.~C.} \bibnamefont{Glotzer}} \bibnamefont{and}
  \bibinfo{author}{\bibfnamefont{C.}~\bibnamefont{Donati}},
  \bibinfo{journal}{J. Phys.: Condens. Matter} \textbf{\bibinfo{volume}{11}},
  \bibinfo{pages}{A285} (\bibinfo{year}{1999}).

\bibitem[{\citenamefont{Simon}(1930)}]{Simon:30}
\bibinfo{author}{\bibfnamefont{F.~E.} \bibnamefont{Simon}},
  \bibinfo{journal}{Naturwiss.} \textbf{\bibinfo{volume}{9}},
  \bibinfo{pages}{244} (\bibinfo{year}{1930}).

\bibitem[{\citenamefont{Kauzmann}(1948)}]{Kauzmann:48}
\bibinfo{author}{\bibfnamefont{W.}~\bibnamefont{Kauzmann}},
  \bibinfo{journal}{Chem. Rev.} \textbf{\bibinfo{volume}{43}},
  \bibinfo{pages}{218} (\bibinfo{year}{1948}).

\bibitem[{\citenamefont{Cavagna et~al.}(2003)\citenamefont{Cavagna, Giardina,
  and Grigera}}]{Cavagna:03}
\bibinfo{author}{\bibfnamefont{A.}~\bibnamefont{Cavagna}},
  \bibinfo{author}{\bibfnamefont{I.}~\bibnamefont{Giardina}}, \bibnamefont{and}
  \bibinfo{author}{\bibfnamefont{T.~S.} \bibnamefont{Grigera}},
  \bibinfo{journal}{J. Chem. Phys.} \textbf{\bibinfo{volume}{118}},
  \bibinfo{pages}{6974} (\bibinfo{year}{2003}).

\bibitem[{\citenamefont{Debenedetti et~al.}(1999)\citenamefont{Debenedetti,
  Stllinger, Truskett, and Roberts}}]{Debenedetti:99}
\bibinfo{author}{\bibfnamefont{P.~G.} \bibnamefont{Debenedetti}},
  \bibinfo{author}{\bibfnamefont{F.~H.} \bibnamefont{Stllinger}},
  \bibinfo{author}{\bibfnamefont{T.~M.} \bibnamefont{Truskett}},
  \bibnamefont{and} \bibinfo{author}{\bibfnamefont{C.~J.}
  \bibnamefont{Roberts}}, \bibinfo{journal}{J. Phys. Chem. B}
  \textbf{\bibinfo{volume}{103}}, \bibinfo{pages}{7390} (\bibinfo{year}{1999}).

\bibitem[{\citenamefont{Nave et~al.}(2002)\citenamefont{Nave, Mossa, and
  Sciortino}}]{Nave:02}
\bibinfo{author}{\bibfnamefont{E.~L.} \bibnamefont{Nave}},
  \bibinfo{author}{\bibfnamefont{S.}~\bibnamefont{Mossa}}, \bibnamefont{and}
  \bibinfo{author}{\bibfnamefont{F.}~\bibnamefont{Sciortino}},
  \bibinfo{journal}{Phys. Rev. Lett.} \textbf{\bibinfo{volume}{88}},
  \bibinfo{pages}{225701} (\bibinfo{year}{2002}).

\bibitem[{\citenamefont{Shell et~al.}(2003)\citenamefont{Shell, Debenedetti,
  Nave, and Sciortino}}]{Shell:03}
\bibinfo{author}{\bibfnamefont{M.~S.} \bibnamefont{Shell}},
  \bibinfo{author}{\bibfnamefont{P.~G.} \bibnamefont{Debenedetti}},
  \bibinfo{author}{\bibfnamefont{E.~L.} \bibnamefont{Nave}}, \bibnamefont{and}
  \bibinfo{author}{\bibfnamefont{F.}~\bibnamefont{Sciortino}},
  \bibinfo{journal}{J. Chem. Phys.} \textbf{\bibinfo{volume}{118}},
  \bibinfo{pages}{8821} (\bibinfo{year}{2003}).

\bibitem[{\citenamefont{Gibbs and Dimarzio}(1958)}]{Gibbs:58}
\bibinfo{author}{\bibfnamefont{J.~H.} \bibnamefont{Gibbs}} \bibnamefont{and}
  \bibinfo{author}{\bibfnamefont{E.~A.} \bibnamefont{Dimarzio}},
  \bibinfo{journal}{J. Chem. Phys.} \textbf{\bibinfo{volume}{28}},
  \bibinfo{pages}{373} (\bibinfo{year}{1958}).

\bibitem[{\citenamefont{Angell}(1997)}]{Angell:97}
\bibinfo{author}{\bibfnamefont{C.~A.} \bibnamefont{Angell}},
  \bibinfo{journal}{J. Res. NIST} \textbf{\bibinfo{volume}{102}},
  \bibinfo{pages}{171} (\bibinfo{year}{1997}).

\bibitem[{\citenamefont{Stillinger}(1988)}]{Stillinger:88}
\bibinfo{author}{\bibfnamefont{F.~H.} \bibnamefont{Stillinger}},
  \bibinfo{journal}{J. Chem. Phys.} \textbf{\bibinfo{volume}{88}},
  \bibinfo{pages}{7818} (\bibinfo{year}{1988}).

\bibitem[{\citenamefont{Garrahan and Chandler}(2002)}]{Garrahan:02}
\bibinfo{author}{\bibfnamefont{J.~P.} \bibnamefont{Garrahan}} \bibnamefont{and}
  \bibinfo{author}{\bibfnamefont{D.}~\bibnamefont{Chandler}},
  \bibinfo{journal}{Phys. Rev. Lett.} \textbf{\bibinfo{volume}{89}},
  \bibinfo{pages}{035704} (\bibinfo{year}{2002}).

\bibitem[{\citenamefont{Garrahan and Chandler}(2003)}]{Garrahan:03}
\bibinfo{author}{\bibfnamefont{J.~P.} \bibnamefont{Garrahan}} \bibnamefont{and}
  \bibinfo{author}{\bibfnamefont{D.}~\bibnamefont{Chandler}},
  \bibinfo{journal}{Proc. Natl. Acad. Sci. USA} \textbf{\bibinfo{volume}{100}},
  \bibinfo{pages}{9710} (\bibinfo{year}{2003}).

\bibitem[{\citenamefont{Macedo et~al.}(1966)\citenamefont{Macedo, Capps, and
  Litovitz}}]{Macedo:66}
\bibinfo{author}{\bibfnamefont{P.~B.} \bibnamefont{Macedo}},
  \bibinfo{author}{\bibfnamefont{W.}~\bibnamefont{Capps}}, \bibnamefont{and}
  \bibinfo{author}{\bibfnamefont{T.~A.} \bibnamefont{Litovitz}},
  \bibinfo{journal}{J. Chem. Phys.} \textbf{\bibinfo{volume}{44}},
  \bibinfo{pages}{3357} (\bibinfo{year}{1966}).

\bibitem[{\citenamefont{Angell and Rao}(1972)}]{Angell:72}
\bibinfo{author}{\bibfnamefont{C.~A.} \bibnamefont{Angell}} \bibnamefont{and}
  \bibinfo{author}{\bibfnamefont{K.~J.} \bibnamefont{Rao}},
  \bibinfo{journal}{J. Chem. Phys.} \textbf{\bibinfo{volume}{57}},
  \bibinfo{pages}{470} (\bibinfo{year}{1972}).

\bibitem[{\citenamefont{Perez}(1985)}]{Perez:85}
\bibinfo{author}{\bibfnamefont{J.}~\bibnamefont{Perez}}, \bibinfo{journal}{J.
  Phys.} \textbf{\bibinfo{volume}{C10}}, \bibinfo{pages}{427}
  (\bibinfo{year}{1985}).

\bibitem[{\citenamefont{Angell}(2000)}]{Angell:00}
\bibinfo{author}{\bibfnamefont{C.~A.} \bibnamefont{Angell}},
  \bibinfo{journal}{J. Phys.: Condens. Matter} \textbf{\bibinfo{volume}{12}},
  \bibinfo{pages}{6463} (\bibinfo{year}{2000}).

\bibitem[{\citenamefont{Moynihan and Angell}(2000)}]{Moynihan:00}
\bibinfo{author}{\bibfnamefont{C.~T.} \bibnamefont{Moynihan}} \bibnamefont{and}
  \bibinfo{author}{\bibfnamefont{C.~A.} \bibnamefont{Angell}},
  \bibinfo{journal}{J. Non-Crystal. Sol.} \textbf{\bibinfo{volume}{274}},
  \bibinfo{pages}{131} (\bibinfo{year}{2000}).

\bibitem[{\citenamefont{Angell et~al.}(1977)\citenamefont{Angell, Williams,
  Rao, and Tucker}}]{Angell:77}
\bibinfo{author}{\bibfnamefont{C.~A.} \bibnamefont{Angell}},
  \bibinfo{author}{\bibfnamefont{E.}~\bibnamefont{Williams}},
  \bibinfo{author}{\bibfnamefont{K.~J.} \bibnamefont{Rao}}, \bibnamefont{and}
  \bibinfo{author}{\bibfnamefont{J.~C.} \bibnamefont{Tucker}},
  \bibinfo{journal}{J. Chem. Phys.} \textbf{\bibinfo{volume}{81}},
  \bibinfo{pages}{238} (\bibinfo{year}{1977}).

\bibitem[{\citenamefont{Hemmati et~al.}(2001)\citenamefont{Hemmati, Moynihan,
  and Angell}}]{Hemmati:01}
\bibinfo{author}{\bibfnamefont{M.}~\bibnamefont{Hemmati}},
  \bibinfo{author}{\bibfnamefont{C.~T.} \bibnamefont{Moynihan}},
  \bibnamefont{and} \bibinfo{author}{\bibfnamefont{C.~A.}
  \bibnamefont{Angell}}, \bibinfo{journal}{J. Chem. Phys.}
  \textbf{\bibinfo{volume}{115}}, \bibinfo{pages}{6663} (\bibinfo{year}{2001}).

\bibitem[{\citenamefont{Debenedetti et~al.}(2003)\citenamefont{Debenedetti,
  Stillinger, and Shell}}]{Debenedetti:03}
\bibinfo{author}{\bibfnamefont{P.~G.} \bibnamefont{Debenedetti}},
  \bibinfo{author}{\bibfnamefont{F.~H.} \bibnamefont{Stillinger}},
  \bibnamefont{and} \bibinfo{author}{\bibfnamefont{M.~S.} \bibnamefont{Shell}},
  \bibinfo{journal}{J. Phys. Chem. B} \textbf{\bibinfo{volume}{107}},
  \bibinfo{pages}{14434} (\bibinfo{year}{2003}).

\bibitem[{\citenamefont{Tanaka}(1998)}]{Tanaka:98}
\bibinfo{author}{\bibfnamefont{H.}~\bibnamefont{Tanaka}}, \bibinfo{journal}{J.
  Phys.: Condens. Matter} \textbf{\bibinfo{volume}{10}}, \bibinfo{pages}{L207}
  (\bibinfo{year}{1998}).

\bibitem[{\citenamefont{Tanaka}(1999)}]{Tanaka1:99}
\bibinfo{author}{\bibfnamefont{H.}~\bibnamefont{Tanaka}}, \bibinfo{journal}{J.
  Chem. Phys.} \textbf{\bibinfo{volume}{111}}, \bibinfo{pages}{3163}
  (\bibinfo{year}{1999}), \bibinfo{note}{\textbf{111}, 3175 (1999).}

\bibitem[{\citenamefont{Granato}(1992)}]{Granato:92}
\bibinfo{author}{\bibfnamefont{A.~V.} \bibnamefont{Granato}},
  \bibinfo{journal}{Phys. Rev. Lett.} \textbf{\bibinfo{volume}{68}},
  \bibinfo{pages}{974} (\bibinfo{year}{1992}).

\bibitem[{\citenamefont{Granato}(2002)}]{Granato:02}
\bibinfo{author}{\bibfnamefont{A.~V.} \bibnamefont{Granato}},
  \bibinfo{journal}{J. Non-Cryst. Solids} \textbf{\bibinfo{volume}{307-310}},
  \bibinfo{pages}{376} (\bibinfo{year}{2002}).

\bibitem[{\citenamefont{Xia and Wolynes}(2000)}]{Xia:00}
\bibinfo{author}{\bibfnamefont{X.}~\bibnamefont{Xia}} \bibnamefont{and}
  \bibinfo{author}{\bibfnamefont{P.~G.} \bibnamefont{Wolynes}},
  \bibinfo{journal}{Proc. Nat. Acad. Sci.} \textbf{\bibinfo{volume}{97}},
  \bibinfo{pages}{2990} (\bibinfo{year}{2000}).

\bibitem[{\citenamefont{Lubchenko and Wolynes}(2004)}]{Lubchenko:04}
\bibinfo{author}{\bibfnamefont{V.}~\bibnamefont{Lubchenko}} \bibnamefont{and}
  \bibinfo{author}{\bibfnamefont{P.~G.} \bibnamefont{Wolynes}},
  \bibinfo{journal}{J. Chem. Phys.} \textbf{\bibinfo{volume}{121}},
  \bibinfo{pages}{2852} (\bibinfo{year}{2004}).

\bibitem[{\citenamefont{Derrida}(1980)}]{Derrida:80}
\bibinfo{author}{\bibfnamefont{B.}~\bibnamefont{Derrida}},
  \bibinfo{journal}{Phys. Rev. Lett.} \textbf{\bibinfo{volume}{45}},
  \bibinfo{pages}{79} (\bibinfo{year}{1980}).

\bibitem[{\citenamefont{Derrida}(1981)}]{Derrida:81}
\bibinfo{author}{\bibfnamefont{B.}~\bibnamefont{Derrida}},
  \bibinfo{journal}{Phys. Rev. B} \textbf{\bibinfo{volume}{24}},
  \bibinfo{pages}{2613} (\bibinfo{year}{1981}).

\bibitem[{\citenamefont{Richert and B{\"a}ssler}(1990)}]{Richert:90}
\bibinfo{author}{\bibfnamefont{R.}~\bibnamefont{Richert}} \bibnamefont{and}
  \bibinfo{author}{\bibfnamefont{H.}~\bibnamefont{B{\"a}ssler}},
  \bibinfo{journal}{J. Phys.: Condens. Matter} \textbf{\bibinfo{volume}{2}},
  \bibinfo{pages}{2273} (\bibinfo{year}{1990}).

\bibitem[{\citenamefont{Swallen et~al.}(2003)\citenamefont{Swallen, Bonvallet,
  McMahon, and Ediger}}]{Swallen:03}
\bibinfo{author}{\bibfnamefont{S.~F.} \bibnamefont{Swallen}},
  \bibinfo{author}{\bibfnamefont{P.~A.} \bibnamefont{Bonvallet}},
  \bibinfo{author}{\bibfnamefont{R.~J.} \bibnamefont{McMahon}},
  \bibnamefont{and} \bibinfo{author}{\bibfnamefont{M.~D.}
  \bibnamefont{Ediger}}, \bibinfo{journal}{Phys. Rev. Lett.}
  \textbf{\bibinfo{volume}{90}}, \bibinfo{pages}{015901}
  (\bibinfo{year}{2003}).

\bibitem[{\citenamefont{Speedy and Debenedetti}(1988)}]{Speedy:88}
\bibinfo{author}{\bibfnamefont{R.~J.} \bibnamefont{Speedy}} \bibnamefont{and}
  \bibinfo{author}{\bibfnamefont{P.~G.} \bibnamefont{Debenedetti}},
  \bibinfo{journal}{Mol. Phys.} \textbf{\bibinfo{volume}{88}},
  \bibinfo{pages}{1293} (\bibinfo{year}{1988}).

\bibitem[{\citenamefont{B{\"u}chner and Heuer}(1999)}]{Buchner:99}
\bibinfo{author}{\bibfnamefont{S.}~\bibnamefont{B{\"u}chner}} \bibnamefont{and}
  \bibinfo{author}{\bibfnamefont{A.}~\bibnamefont{Heuer}},
  \bibinfo{journal}{Phys. Rev. E} \textbf{\bibinfo{volume}{60}},
  \bibinfo{pages}{6507} (\bibinfo{year}{1999}).

\bibitem[{\citenamefont{Sciortino et~al.}(2000)\citenamefont{Sciortino, Kob,
  and Tartaglia}}]{Sciortino:00}
\bibinfo{author}{\bibfnamefont{F.}~\bibnamefont{Sciortino}},
  \bibinfo{author}{\bibfnamefont{W.}~\bibnamefont{Kob}}, \bibnamefont{and}
  \bibinfo{author}{\bibfnamefont{P.}~\bibnamefont{Tartaglia}},
  \bibinfo{journal}{J. Phys.: Condens. Matter} \textbf{\bibinfo{volume}{12}},
  \bibinfo{pages}{6525} (\bibinfo{year}{2000}).

\bibitem[{\citenamefont{Heuer and B{\"u}chner}(2000)}]{Heuer:00}
\bibinfo{author}{\bibfnamefont{A.}~\bibnamefont{Heuer}} \bibnamefont{and}
  \bibinfo{author}{\bibfnamefont{S.}~\bibnamefont{B{\"u}chner}},
  \bibinfo{journal}{J. Phys.: Condens. Matter} \textbf{\bibinfo{volume}{12}},
  \bibinfo{pages}{6535} (\bibinfo{year}{2000}).

\bibitem[{\citenamefont{Sastry}(2001)}]{Sastry:01}
\bibinfo{author}{\bibfnamefont{S.}~\bibnamefont{Sastry}},
  \bibinfo{journal}{Nature} \textbf{\bibinfo{volume}{409}},
  \bibinfo{pages}{164} (\bibinfo{year}{2001}).

\bibitem[{\citenamefont{Yan et~al.}(2004)\citenamefont{Yan, Jain, and
  de~Pablo}}]{Yan:04}
\bibinfo{author}{\bibfnamefont{Q.}~\bibnamefont{Yan}},
  \bibinfo{author}{\bibfnamefont{T.~S.} \bibnamefont{Jain}}, \bibnamefont{and}
  \bibinfo{author}{\bibfnamefont{J.~J.} \bibnamefont{de~Pablo}},
  \bibinfo{journal}{Phys. Rev. Lett.} \textbf{\bibinfo{volume}{92}},
  \bibinfo{pages}{235701} (\bibinfo{year}{2004}).

\bibitem[{\citenamefont{Grigera and Parisi}(2001)}]{Grigera:01}
\bibinfo{author}{\bibfnamefont{T.}~\bibnamefont{Grigera}} \bibnamefont{and}
  \bibinfo{author}{\bibfnamefont{G.}~\bibnamefont{Parisi}},
  \bibinfo{journal}{Phys. Rev. E} \textbf{\bibinfo{volume}{63}}
  (\bibinfo{year}{2001}).

\bibitem[{\citenamefont{Yu and Carruzzo}()}]{Yu:02}
\bibinfo{author}{\bibfnamefont{C.~C.} \bibnamefont{Yu}} \bibnamefont{and}
  \bibinfo{author}{\bibfnamefont{H.~M.} \bibnamefont{Carruzzo}},
  \bibinfo{note}{cond-mat/0209221}.

\bibitem[{\citenamefont{Yu and Carruzzo}(2004)}]{Yu:04}
\bibinfo{author}{\bibfnamefont{C.~C.} \bibnamefont{Yu}} \bibnamefont{and}
  \bibinfo{author}{\bibfnamefont{H.~M.} \bibnamefont{Carruzzo}},
  \bibinfo{journal}{Phys. Rev. E} \textbf{\bibinfo{volume}{69}},
  \bibinfo{pages}{051201} (\bibinfo{year}{2004}).

\bibitem[{\citenamefont{Angell}(1980)}]{Angell:80}
\bibinfo{author}{\bibfnamefont{C.~A.} \bibnamefont{Angell}}, in
  \emph{\bibinfo{booktitle}{Vibrational Spectroscopy in Molecular Liquids and
  Solids}}, edited by \bibinfo{editor}{\bibfnamefont{E.}~\bibnamefont{Pick}}
  \bibnamefont{and} \bibinfo{editor}{\bibfnamefont{S.}~\bibnamefont{Bratos}}
  (\bibinfo{publisher}{Plenum Press}, \bibinfo{year}{1980}), p.
  \bibinfo{pages}{187}.

\bibitem[{\citenamefont{Mossa et~al.}(2002)\citenamefont{Mossa, Nave, Stanley,
  Donati, Sciortino, and Tartaglia}}]{Mossa:02}
\bibinfo{author}{\bibfnamefont{S.}~\bibnamefont{Mossa}},
  \bibinfo{author}{\bibfnamefont{E.~L.} \bibnamefont{Nave}},
  \bibinfo{author}{\bibfnamefont{H.~E.} \bibnamefont{Stanley}},
  \bibinfo{author}{\bibfnamefont{C.}~\bibnamefont{Donati}},
  \bibinfo{author}{\bibfnamefont{F.}~\bibnamefont{Sciortino}},
  \bibnamefont{and}
  \bibinfo{author}{\bibfnamefont{P.}~\bibnamefont{Tartaglia}},
  \bibinfo{journal}{Phys. Rev. E} \textbf{\bibinfo{volume}{65}},
  \bibinfo{pages}{041205} (\bibinfo{year}{2002}).

\bibitem[{\citenamefont{Angell et~al.}(2003)\citenamefont{Angell, Yue, Wang,
  Copley, Borick, and Mossa}}]{Angell:03}
\bibinfo{author}{\bibfnamefont{A.~A.} \bibnamefont{Angell}},
  \bibinfo{author}{\bibfnamefont{Y.}~\bibnamefont{Yue}},
  \bibinfo{author}{\bibfnamefont{L.-M.} \bibnamefont{Wang}},
  \bibinfo{author}{\bibfnamefont{J.~R.~D.} \bibnamefont{Copley}},
  \bibinfo{author}{\bibfnamefont{S.}~\bibnamefont{Borick}}, \bibnamefont{and}
  \bibinfo{author}{\bibfnamefont{S.}~\bibnamefont{Mossa}}, \bibinfo{journal}{J.
  Phys.: Condens. Matter} \textbf{\bibinfo{volume}{15}}, \bibinfo{pages}{1}
  (\bibinfo{year}{2003}).

\bibitem[{\citenamefont{Gurevich et~al.}(2003)\citenamefont{Gurevich, Parshin,
  and Schober}}]{Gurevich:03}
\bibinfo{author}{\bibfnamefont{V.~L.} \bibnamefont{Gurevich}},
  \bibinfo{author}{\bibfnamefont{D.~A.} \bibnamefont{Parshin}},
  \bibnamefont{and} \bibinfo{author}{\bibfnamefont{H.~R.}
  \bibnamefont{Schober}}, \bibinfo{journal}{Phys. Rev. B}
  \textbf{\bibinfo{volume}{67}}, \bibinfo{pages}{094203}
  (\bibinfo{year}{2003}).

\bibitem[{\citenamefont{Angell and Wong}(1970)}]{Angell:70}
\bibinfo{author}{\bibfnamefont{C.~A.} \bibnamefont{Angell}} \bibnamefont{and}
  \bibinfo{author}{\bibfnamefont{J.~J.} \bibnamefont{Wong}},
  \bibinfo{journal}{J. Chem. Phys.} \textbf{\bibinfo{volume}{53}},
  \bibinfo{pages}{2053} (\bibinfo{year}{1970}).

\bibitem[{\citenamefont{Angell}(2004)}]{Angell:04}
\bibinfo{author}{\bibfnamefont{C.~A.} \bibnamefont{Angell}},
  \bibinfo{journal}{J. Phys.: Condens. Matter} \textbf{\bibinfo{volume}{16}},
  \bibinfo{pages}{S5153} (\bibinfo{year}{2004}).

\bibitem[{com({\natexlab{a}})}]{com:1}
\bibinfo{note}{In the case of true networks, the thermodynamics is dominated by
  the breaking of network
  bonds\cite{Sastry:01,Voivod:01,Saksaengwijit:04,Voivod:04} though the
  observed behavior does not conform to the simple random bond breaking
  behavior, the broken bonds tending to cluster together. Even on the short
  time scales of molecular dynamics simulations, a completely bonded state can
  be realized before equilibrium is
  lost.\cite{Sastry:01,Voivod:01,Saksaengwijit:04,Voivod:04} However,
  considerable non-vibrational entropy remains in the amorphous solid,
  associated with the different manner in which the completely bonded network
  can be connected.}

\bibitem[{\citenamefont{Stillinger and Weber}(1982)}]{Stillinger:82}
\bibinfo{author}{\bibfnamefont{F.~H.} \bibnamefont{Stillinger}}
  \bibnamefont{and} \bibinfo{author}{\bibfnamefont{T.~A.} \bibnamefont{Weber}},
  \bibinfo{journal}{Phys. Rev. A} \textbf{\bibinfo{volume}{25}},
  \bibinfo{pages}{978} (\bibinfo{year}{1982}).

\bibitem[{\citenamefont{Bryngelson and Wolynes}(1987)}]{Bryngelson:87}
\bibinfo{author}{\bibfnamefont{J.~D.} \bibnamefont{Bryngelson}}
  \bibnamefont{and} \bibinfo{author}{\bibfnamefont{P.~G.}
  \bibnamefont{Wolynes}}, \bibinfo{journal}{Proc. Natl. Acad. Sci.}
  \textbf{\bibinfo{volume}{84}}, \bibinfo{pages}{7524} (\bibinfo{year}{1987}).

\bibitem[{\citenamefont{Saika-Voivod et~al.}(2001)\citenamefont{Saika-Voivod,
  Poole, and Sciortino}}]{Voivod:01}
\bibinfo{author}{\bibfnamefont{I.}~\bibnamefont{Saika-Voivod}},
  \bibinfo{author}{\bibfnamefont{P.~H.} \bibnamefont{Poole}}, \bibnamefont{and}
  \bibinfo{author}{\bibfnamefont{F.}~\bibnamefont{Sciortino}},
  \bibinfo{journal}{Nature} \textbf{\bibinfo{volume}{412}},
  \bibinfo{pages}{514} (\bibinfo{year}{2001}).

\bibitem[{\citenamefont{Saika-Voivod et~al.}(2004)\citenamefont{Saika-Voivod,
  Sciortino, and Poole}}]{Voivod:04}
\bibinfo{author}{\bibfnamefont{I.}~\bibnamefont{Saika-Voivod}},
  \bibinfo{author}{\bibfnamefont{F.}~\bibnamefont{Sciortino}},
  \bibnamefont{and} \bibinfo{author}{\bibfnamefont{P.~H.} \bibnamefont{Poole}},
  \bibinfo{journal}{Phys. Rev. E} \textbf{\bibinfo{volume}{69}},
  \bibinfo{pages}{041503} (\bibinfo{year}{2004}).

\bibitem[{\citenamefont{Saksaengwijit et~al.}(2004)\citenamefont{Saksaengwijit,
  Reinisch, and Heuer}}]{Saksaengwijit:04}
\bibinfo{author}{\bibfnamefont{A.}~\bibnamefont{Saksaengwijit}},
  \bibinfo{author}{\bibfnamefont{J.}~\bibnamefont{Reinisch}}, \bibnamefont{and}
  \bibinfo{author}{\bibfnamefont{A.}~\bibnamefont{Heuer}},
  \bibinfo{journal}{Phys. Rev. Lett.} \textbf{\bibinfo{volume}{93}}
  (\bibinfo{year}{2004}).

\bibitem[{\citenamefont{Odagaki et~al.}(2002)\citenamefont{Odagaki, Yoshidome,
  Tao, and Yoshimori}}]{Odagaki:02}
\bibinfo{author}{\bibfnamefont{T.}~\bibnamefont{Odagaki}},
  \bibinfo{author}{\bibfnamefont{T.}~\bibnamefont{Yoshidome}},
  \bibinfo{author}{\bibfnamefont{T.}~\bibnamefont{Tao}}, \bibnamefont{and}
  \bibinfo{author}{\bibfnamefont{A.}~\bibnamefont{Yoshimori}},
  \bibinfo{journal}{J. Chem. Phys.} \textbf{\bibinfo{volume}{117}},
  \bibinfo{pages}{10151} (\bibinfo{year}{2002}).

\bibitem[{\citenamefont{Landau and Lifshits}(1980)}]{Landau5}
\bibinfo{author}{\bibfnamefont{L.~D.} \bibnamefont{Landau}} \bibnamefont{and}
  \bibinfo{author}{\bibfnamefont{E.~M.} \bibnamefont{Lifshits}},
  \emph{\bibinfo{title}{Statistical physics}} (\bibinfo{publisher}{Pergamon
  Press}, \bibinfo{address}{New York}, \bibinfo{year}{1980}).

\bibitem[{\citenamefont{B{\"a}ssler}(1987)}]{Baessler:87}
\bibinfo{author}{\bibfnamefont{H.}~\bibnamefont{B{\"a}ssler}},
  \bibinfo{journal}{Phys. Rev. Lett.} \textbf{\bibinfo{volume}{58}},
  \bibinfo{pages}{767} (\bibinfo{year}{1987}).

\bibitem[{\citenamefont{Marcus}(1993)}]{Marcus:93}
\bibinfo{author}{\bibfnamefont{R.~A.} \bibnamefont{Marcus}},
  \bibinfo{journal}{Rev. Mod. Phys.} \textbf{\bibinfo{volume}{65}},
  \bibinfo{pages}{599} (\bibinfo{year}{1993}).

\bibitem[{\citenamefont{Privalko}(1980)}]{Privalko:80}
\bibinfo{author}{\bibfnamefont{Y.}~\bibnamefont{Privalko}},
  \bibinfo{journal}{J. Phys. Chem.} \textbf{\bibinfo{volume}{84}},
  \bibinfo{pages}{3307} (\bibinfo{year}{1980}).

\bibitem[{\citenamefont{Aba et~al.}(1990)\citenamefont{Aba, Busse, List, and
  Angell}}]{Aba:90}
\bibinfo{author}{\bibfnamefont{C.}~\bibnamefont{Aba}},
  \bibinfo{author}{\bibfnamefont{L.~E.} \bibnamefont{Busse}},
  \bibinfo{author}{\bibfnamefont{D.~J.} \bibnamefont{List}}, \bibnamefont{and}
  \bibinfo{author}{\bibfnamefont{C.~A.} \bibnamefont{Angell}},
  \bibinfo{journal}{J. Chem. Phys.} \textbf{\bibinfo{volume}{92}},
  \bibinfo{pages}{617} (\bibinfo{year}{1990}).

\bibitem[{\citenamefont{Sastry et~al.}(1998)\citenamefont{Sastry, Debenedetti,
  and Stllinger}}]{Sastry:98}
\bibinfo{author}{\bibfnamefont{S.}~\bibnamefont{Sastry}},
  \bibinfo{author}{\bibfnamefont{P.~G.} \bibnamefont{Debenedetti}},
  \bibnamefont{and} \bibinfo{author}{\bibfnamefont{F.~H.}
  \bibnamefont{Stllinger}}, \bibinfo{journal}{Nature}
  \textbf{\bibinfo{volume}{393}}, \bibinfo{pages}{554} (\bibinfo{year}{1998}).

\bibitem[{\citenamefont{Doliwa and Heuer}(2003)}]{Doliwa:03}
\bibinfo{author}{\bibfnamefont{B.}~\bibnamefont{Doliwa}} \bibnamefont{and}
  \bibinfo{author}{\bibfnamefont{A.}~\bibnamefont{Heuer}},
  \bibinfo{journal}{Phys. Rev. E} \textbf{\bibinfo{volume}{67}},
  \bibinfo{pages}{031506} (\bibinfo{year}{2003}).

\bibitem[{\citenamefont{Chowdhary and Keyes}(2004)}]{Chowdhary:04}
\bibinfo{author}{\bibfnamefont{J.}~\bibnamefont{Chowdhary}} \bibnamefont{and}
  \bibinfo{author}{\bibfnamefont{T.}~\bibnamefont{Keyes}}, \bibinfo{journal}{J.
  Phys. Chem. B} \textbf{\bibinfo{volume}{108}}, \bibinfo{pages}{19786}
  (\bibinfo{year}{2004}).

\bibitem[{\citenamefont{Feynman and F.~L.~Vernon}(1963)}]{Feynman:63}
\bibinfo{author}{\bibfnamefont{R.~P.} \bibnamefont{Feynman}} \bibnamefont{and}
  \bibinfo{author}{\bibfnamefont{J.}~\bibnamefont{F.~L.~Vernon}},
  \bibinfo{journal}{Ann. Phys.} \textbf{\bibinfo{volume}{24}},
  \bibinfo{pages}{118} (\bibinfo{year}{1963}).

\bibitem[{\citenamefont{Sastry and Angell}(2003)}]{Sastry:03}
\bibinfo{author}{\bibfnamefont{S.}~\bibnamefont{Sastry}} \bibnamefont{and}
  \bibinfo{author}{\bibfnamefont{C.~A.} \bibnamefont{Angell}},
  \bibinfo{journal}{Nature Materials} \textbf{\bibinfo{volume}{2}},
  \bibinfo{pages}{739} (\bibinfo{year}{2003}).

\bibitem[{\citenamefont{Tanaka et~al.}(2004)\citenamefont{Tanaka, Kurita, and
  Mataki}}]{Tanaka:04}
\bibinfo{author}{\bibfnamefont{H.}~\bibnamefont{Tanaka}},
  \bibinfo{author}{\bibfnamefont{R.}~\bibnamefont{Kurita}}, \bibnamefont{and}
  \bibinfo{author}{\bibfnamefont{H.}~\bibnamefont{Mataki}},
  \bibinfo{journal}{Phys. Rev. Lett.} \textbf{\bibinfo{volume}{92}},
  \bibinfo{pages}{025701} (\bibinfo{year}{2004}).

\bibitem[{\citenamefont{Denny et~al.}(2003)\citenamefont{Denny, Reichman, and
  Bouchaud}}]{Denny:03}
\bibinfo{author}{\bibfnamefont{R.~A.} \bibnamefont{Denny}},
  \bibinfo{author}{\bibfnamefont{D.~R.} \bibnamefont{Reichman}},
  \bibnamefont{and} \bibinfo{author}{\bibfnamefont{J.-P.}
  \bibnamefont{Bouchaud}}, \bibinfo{journal}{Phys. Rev. Lett.}
  \textbf{\bibinfo{volume}{90}}, \bibinfo{pages}{025503}
  (\bibinfo{year}{2003}).

\bibitem[{com({\natexlab{b}})}]{com:2}
\bibinfo{note}{The results presented in Fig.\ \ref{fig:5} have been extracted
  from Fig.\ 2 of Ref.\ \onlinecite{Denny:03}. For temperature $T^*=0.669$, the
  lowest-energy point substantially deviating from the fit was dropped.}

\bibitem[{\citenamefont{Weber and Stillinger}(1985)}]{Weber:85}
\bibinfo{author}{\bibfnamefont{T.~A.} \bibnamefont{Weber}} \bibnamefont{and}
  \bibinfo{author}{\bibfnamefont{F.~H.} \bibnamefont{Stillinger}},
  \bibinfo{journal}{Phys. Rev. B} \textbf{\bibinfo{volume}{31}},
  \bibinfo{pages}{1954} (\bibinfo{year}{1985}).

\bibitem[{\citenamefont{Kob and Andersen}(1995)}]{Kob:95}
\bibinfo{author}{\bibfnamefont{W.}~\bibnamefont{Kob}} \bibnamefont{and}
  \bibinfo{author}{\bibfnamefont{H.~C.} \bibnamefont{Andersen}},
  \bibinfo{journal}{Phys. Rev. E} \textbf{\bibinfo{volume}{51}},
  \bibinfo{pages}{4626} (\bibinfo{year}{1995}).

\bibitem[{\citenamefont{Stillinger}(1995)}]{Stillinger:95}
\bibinfo{author}{\bibfnamefont{F.~H.} \bibnamefont{Stillinger}},
  \bibinfo{journal}{Science} \textbf{\bibinfo{volume}{267}},
  \bibinfo{pages}{1935} (\bibinfo{year}{1995}).

\bibitem[{\citenamefont{Sciortino et~al.}(1999)\citenamefont{Sciortino, Kob,
  and Tartaglia}}]{SciortinoPRL:99}
\bibinfo{author}{\bibfnamefont{F.}~\bibnamefont{Sciortino}},
  \bibinfo{author}{\bibfnamefont{W.}~\bibnamefont{Kob}}, \bibnamefont{and}
  \bibinfo{author}{\bibfnamefont{P.}~\bibnamefont{Tartaglia}},
  \bibinfo{journal}{Phys. Rev. Lett.} \textbf{\bibinfo{volume}{83}},
  \bibinfo{pages}{3214} (\bibinfo{year}{1999}).

\bibitem[{\citenamefont{Angell}(1995)}]{Angell:95}
\bibinfo{author}{\bibfnamefont{C.~A.} \bibnamefont{Angell}},
  \bibinfo{journal}{Science} \textbf{\bibinfo{volume}{267}},
  \bibinfo{pages}{1924} (\bibinfo{year}{1995}), \bibinfo{note}{see Fig.\ 1}.

\bibitem[{\citenamefont{Busch}(2000)}]{Busch:00}
\bibinfo{author}{\bibfnamefont{R.}~\bibnamefont{Busch}}, \bibinfo{journal}{J.
  of Metals} \textbf{\bibinfo{volume}{52}}, \bibinfo{pages}{39}
  (\bibinfo{year}{2000}).

\bibitem[{\citenamefont{Goldstein}(1976)}]{Goldstein:76}
\bibinfo{author}{\bibfnamefont{M.~J.} \bibnamefont{Goldstein}},
  \bibinfo{journal}{J. Chem. Phys.} \textbf{\bibinfo{volume}{64}},
  \bibinfo{pages}{4767} (\bibinfo{year}{1976}).

\bibitem[{\citenamefont{Richert}(2000)}]{Ranko:00}
\bibinfo{author}{\bibfnamefont{R.}~\bibnamefont{Richert}}, \bibinfo{journal}{J.
  Chem. Phys.} \textbf{\bibinfo{volume}{113}}, \bibinfo{pages}{8404}
  (\bibinfo{year}{2000}).

\bibitem[{\citenamefont{Vath et~al.}(1999)\citenamefont{Vath, Zimmt, Matyushov,
  and Voth}}]{DMjpcb:99}
\bibinfo{author}{\bibfnamefont{P.}~\bibnamefont{Vath}},
  \bibinfo{author}{\bibfnamefont{M.~B.} \bibnamefont{Zimmt}},
  \bibinfo{author}{\bibfnamefont{D.~V.} \bibnamefont{Matyushov}},
  \bibnamefont{and} \bibinfo{author}{\bibfnamefont{G.~A.} \bibnamefont{Voth}},
  \bibinfo{journal}{J. Phys. Chem. B} \textbf{\bibinfo{volume}{103}},
  \bibinfo{pages}{9130} (\bibinfo{year}{1999}).

\bibitem[{\citenamefont{Takeda et~al.}(1999)\citenamefont{Takeda, Yamamuro,
  Tsukushi, Matsuo, and Suga}}]{Takeda:99}
\bibinfo{author}{\bibfnamefont{K.}~\bibnamefont{Takeda}},
  \bibinfo{author}{\bibfnamefont{O.}~\bibnamefont{Yamamuro}},
  \bibinfo{author}{\bibfnamefont{I.}~\bibnamefont{Tsukushi}},
  \bibinfo{author}{\bibfnamefont{T.}~\bibnamefont{Matsuo}}, \bibnamefont{and}
  \bibinfo{author}{\bibfnamefont{H.}~\bibnamefont{Suga}}, \bibinfo{journal}{J.
  Molec. Struct.} \textbf{\bibinfo{volume}{479}}, \bibinfo{pages}{227}
  (\bibinfo{year}{1999}).

\bibitem[{\citenamefont{Dasgupta and Valls}(1999)}]{Dasgupta:99}
\bibinfo{author}{\bibfnamefont{C.}~\bibnamefont{Dasgupta}} \bibnamefont{and}
  \bibinfo{author}{\bibfnamefont{O.~T.} \bibnamefont{Valls}},
  \bibinfo{journal}{Phys. Rev. E} \textbf{\bibinfo{volume}{59}},
  \bibinfo{pages}{3123} (\bibinfo{year}{1999}).

\bibitem[{\citenamefont{Angell et~al.}(1999)\citenamefont{Angell, Richards, and
  Velikov}}]{AngellRichards:99}
\bibinfo{author}{\bibfnamefont{C.~A.} \bibnamefont{Angell}},
  \bibinfo{author}{\bibfnamefont{B.~E.} \bibnamefont{Richards}},
  \bibnamefont{and} \bibinfo{author}{\bibfnamefont{V.}~\bibnamefont{Velikov}},
  \bibinfo{journal}{J. Phys.: Condens. Matter} \textbf{\bibinfo{volume}{11}},
  \bibinfo{pages}{A75} (\bibinfo{year}{1999}).

\bibitem[{\citenamefont{Logan}(1987)}]{Logan:87}
\bibinfo{author}{\bibfnamefont{D.~E.} \bibnamefont{Logan}},
  \bibinfo{journal}{J. Chem. Phys.} \textbf{\bibinfo{volume}{86}},
  \bibinfo{pages}{234} (\bibinfo{year}{1987}).

\bibitem[{\citenamefont{Matyushov and Okhrimovskyy}(2005)}]{DMjcp3:05}
\bibinfo{author}{\bibfnamefont{D.~V.} \bibnamefont{Matyushov}}
  \bibnamefont{and}
  \bibinfo{author}{\bibfnamefont{A.}~\bibnamefont{Okhrimovskyy}},
  \bibinfo{journal}{J. Chem. Phys.} \textbf{\bibinfo{volume}{122}},
  \bibinfo{pages}{in press} (\bibinfo{year}{2005}).

\end{thebibliography}

\end{document}